\documentclass[twocolumn]{aastex631}

\usepackage{amsmath}
\usepackage{xcolor}

\newcommand{\gaia}{{\em Gaia}}
\newcommand{\kms}{km~s$^{-1}$}

\shorttitle{SMC GMM}
\shortauthors{Almeida et al.}

\begin{document}

\title{Population Mixture Modeling of the Small Magellanic Cloud}

\author[0009-0000-0733-2479]{Andr\'es Almeida}
\affiliation{Department of Astronomy, University of Virginia, 530 McCormick Rd, Charlottesville, VA 22904, USA}
\correspondingauthor{Andr\'es Almeida}
\email{tac6na@virginia.edu}

\author[0000-0003-2025-3147]{Steven R. Majewski}
\affiliation{Department of Astronomy, University of Virginia, 530 McCormick Rd, Charlottesville, VA 22904, USA}

\author[0000-0001-5388-0994]{Sten Hasselquist}
\affiliation{Space Telescope Science Institute, 3700 San Martin Drive, Baltimore, MD 21218, USA}

\author[0000-0002-7743-8129]{Claire E. Murray}
\affiliation{Space Telescope Science Institute, 3700 San Martin Drive, Baltimore, MD 21218, USA}

\author[0009-0000-9037-1697]{Slater Oden}
\affiliation{Department of Physics, Montana State University, P.O. Box 173840, Bozeman, MT 59717-3840, USA}

\author[0000-0002-1793-3689]{David L. Nidever}
\affiliation{Department of Physics, Montana State University, P.O. Box 173840, Bozeman, MT 59717-3840, USA}

\author[0000-0002-0572-8012]{Vedant Chandra}
\altaffiliation{NASA Hubble Fellow}
\affiliation{MIT Kavli Institute for Astrophysics and Space Research, 70 Vassar St, Cambridge, MA 02139, USA}
\affiliation{Center for Astrophysics $\mid$ Harvard \& Smithsonian, 60 Garden St, Cambridge, MA 02138, USA}

\author[0000-0002-6797-696X]{Maria-Rosa L. Cioni}
\affiliation{Leibniz-Institut f\"ur Astrophysik Potsdam, An der Sternwarte 16, D-14482 Potsdam, Germany}

\author[0000-0002-4669-9967]{Xiaowei Ou}
\affiliation{Department of Astronomy, University of Virginia, 530 McCormick Rd, Charlottesville, VA 22904, USA}
\affiliation{The NSF-Simons AI Institute for Cosmic Origins, USA}

\author[0000-0003-4254-7111]{Bruno Dias}
\affiliation{Instituto de Astrof{\'\i}sica, Departamento de F{\'\i}sica y Astronom{\'\i}a, Facultad de Ciencias Exactas, Universidad Andr\'es Bello, Av. Fern\'andez Concha 700, Las Condes, Santiago, Chile}

\author[0000-0003-1856-2151]{Danny Horta}
\affiliation{Institute for Astronomy, University of Edinburgh, Royal Observatory, Blackford Hill, Edinburgh, EH9 3HJ, UK}

\author[0000-0002-7955-7359]{Dhanesh Krishnarao}
\affiliation{Department of Physics, Colorado College, 14 East Cache la Poudre St., Colorado Springs, CO, 80903, USA}

\begin{abstract}
The Small Magellanic Cloud (SMC), strongly disturbed by its interaction with the Large Magellanic Cloud (LMC), has a large line-of-sight depth that has often been interpreted as the manifestation of a single, highly inclined body.
We test this hypothesis with 8,683 red giants from the SDSS-V Magellanic Genesis Survey, combining line-of-sight velocities and APOGEE-calibrated metallicities from optical spectra with \gaia\ DR3 proper motions.
A Gaussian mixture model fitted to these observables alone, accounting for each star's uncertainties, favors four components with distinct, overlapping spatial distributions; we characterize them with distances from the tip of the red giant branch, kinematics, APOGEE abundances, and isochrone fits.
A Foreground Component lies at a distance of $50.3\pm1.0$~kpc, in front of the inner body, as a thin sheet.
It has the metallicity and red giant branch of the inner SMC but moves at least 33~\kms\ relative to SMC Halo Candidate stars at the same positions, consistent with inner-SMC material pulled out on the LMC-facing side.
The inner body divides into western and eastern components with similar metallicities, distances, and red giant branches but different mean motions, i.e., either two overlapping populations or one body in two kinematical states.
Both have line-of-sight velocity gradients pointing away from the LMC, and neither shows detectable sky-plane rotation ($\lesssim$5--8~\kms).
A Halo Candidate, the most metal-poor, dynamically hottest, most extended, and oldest component, surrounds the others.
Part of the SMC's depth is thus attributable to a superposition of populations in different kinematical states.
\end{abstract}


\vspace{-3\baselineskip}
\section{Introduction}
\label{sec:intro}

Dwarf--dwarf galaxy encounters and dwarf--host interactions are among the most efficient ways to reshape low-mass galaxies.
Tidal forcing can heat stellar orbits \citep{mayer2001,lokas2010,kazantzidis2011}, drive long-lived distortions \citep{besla2007,putman2003,diaz2012}, and redistribute gas and stars far beyond a galaxy's main body, while star formation and chemical enrichment respond to the same gravitational perturbations.
The Large Magellanic Cloud (LMC) and the Small Magellanic Cloud (SMC) are the nearest interacting pair of dwarf galaxies, and a laboratory in which the above processes can be studied star by star, with three-dimensional velocities from \gaia\ proper motions and spectroscopic line-of-sight velocities for large samples of stars \citep{almeida2024,zivick2021}.
The spectacular, multi-phase tidal features of the Magellanic System---including the Magellanic Stream, the Bridge, and the Leading Arm---extend over many tens of degrees on the sky and demonstrate that its recent interaction history is dynamically violent and still unfolding \citep{besla2012,diaz2012,nidever2013}.
A central goal of Magellanic studies is to connect these large-scale debris features to the internal dynamical state of each Cloud: which stellar populations are being stripped, from which regions, and by what sequence of encounters.

The SMC is particularly compelling in this context.
The SMC's red giants are supported mainly by random motions rather than rotation: estimates of their ordered rotation range from moderate \citep{dobbie2014,zivick2021} to small compared with their velocity dispersion \citep{rathore2025}.
The SMC has a complex morphology (a bar, a wing, and an extended low-surface-brightness periphery), a large line-of-sight depth, and strong asymmetries, all of which are widely interpreted as signatures of repeated tidal stirring \citep[e.g.,][]{crowl2001,subramanian2012,dias2016,pieres2017,massana2020,gaiaMagellanic2021}.
In the \gaia\ era, proper-motion fields have revealed coherent streaming motions and substantial shear across the main body and the outskirts, consistent with tidal stretching and heating rather than with a cold, steadily rotating disk \citep{zivick2018,zivick2021}.
Proper motions from the VISTA survey of the Magellanic Clouds system (VMC; \citealt{cioni2011}) also reveal different flows among stellar populations, with stronger southeastward streaming in red clump and young stars than in older red giants \citep{vijayasree2026}.
Line-of-sight kinematics and chemistry provide complementary diagnostics, probing the dynamical temperature and any embedded rotation versus the imprint of chemical evolution across spatial substructures.
A key challenge is that these observables---line-of-sight velocity, proper motion, and metallicity---vary together and overlap strongly between stellar populations, so physically distinct components are difficult to isolate without a statistical, multivariate approach.

Early spectroscopic and photometric work established foundational, but spatially sparse, constraints on the SMC's kinematical and metallicity distribution.
Radial-velocity surveys of giants and other tracers mapped broad gradients and hinted at multiple kinematical components, but were often limited by heterogeneous target selection and incomplete coverage of the periphery \citep[e.g.,][]{harris2006,evans2008,parisi2009,dobbie2014,james2021}.
More recent wide-area efforts, however, have sharpened the picture, reporting population-dependent trends and metallicity gradients across the bar, the wing, and the outer regions \citep[e.g.,][]{dias2016,carrera2017,parisi2022,debortoli2022,dias2022,elyoussoufi2023,omkumar2025,povick2025}.
The physical interpretation of these trends nevertheless remains debated: do they reflect a single disturbed stellar body, a superposition of tidal debris and a bound remnant, or distinct structural components (e.g., a bar and wing versus a spheroid or halo) with different enrichment and dynamical histories?

Structure along the line of sight adds a further, crucial dimension to the problem.
Red clump (RC) analyses have revealed a large line-of-sight depth and evidence for multiple distance layers in parts of the SMC \citep{subramanian2012,tatton2021}, and RC stars have since been used to construct contiguous distance maps of the entire Magellanic System \citep{oden2025}.
Building on this, recent work has shown that the SMC periphery can exhibit a striking distance bimodality aligned with the direction of the Magellanic Bridge \citep{nidever2013,subramanian2017,el-youssoufi2021,dias2021,james2021,omkumar2021}, which is naturally interpreted in terms of tidal reconfiguration and stripping.
From young massive stars matched to high-resolution H\,{\sc i} observations, \citet{murray2024} further find that the SMC's interstellar medium is arranged into two superimposed star-forming systems separated by $\sim$5~kpc along the line of sight, which suggests that at least part of the SMC's present-day complexity reflects tidal separation into distinct line-of-sight layers.
In our previous work \citep{almeida2024}, APOGEE spectroscopy in eleven fields showed that the nearer stars on the eastern side approach us relative to the rest of the SMC and resemble the central SMC in their metallicity distribution, whereas the more distant stars there resemble the western periphery.
We interpreted the nearer stars as material pulled out of the central SMC.
That analysis relied on sparse fields and regions defined in advance.
Placing this foreground structure within the kinematical and chemical structure of the whole SMC requires contiguous coverage and a joint analysis of velocities, proper motions, and metallicities that takes their measurement uncertainties into account.

Several previous studies have focused on particular aspects of the SMC, including kinematics without chemistry \citep{zivick2021,niederhofer2021}, metallicity distributions without a kinematical analysis \citep{omkumar2025}, and line-of-sight distances without full phase-space information \citep{tatton2021}.
The VISCACHA studies of \citet{dias2021,dias2022} and \citet{parisi2024} combined full phase-space information with ages and metallicities for 24 SMC star clusters.
However, for field stars, the observables can be analyzed jointly, letting the data define the dominant chemo-dynamical groupings while explicitly accounting for each star's measurement uncertainties, which can otherwise blur real structure or create spurious structure.
Gaussian mixture modeling is well suited to this problem, especially for a large sample of stars with line-of-sight velocities, proper motions, and metallicities.
It can describe the distribution of stars in a multi-dimensional space of observables as a sum of Gaussian components and give every star a probability of belonging to each component.
Populations that overlap in every individual observable can then be separated by their joint behavior, without imposing cuts between components in advance, and the spatial distribution, kinematics, and chemistry of each component can be examined in turn.

In this work, we use a homogeneous sample of 8,683 red giant branch (RGB) stars with line-of-sight velocities and metallicities from optical spectroscopy obtained by the Magellanic Genesis Survey of the Sloan Digital Sky Survey V (SDSS-V; \citealt{Nidever2026}), combined with \gaia\ DR3 proper motions \citep{gaiadr3}.
We fit a Gaussian mixture model to the joint distribution of line-of-sight velocity, proper motion, and [Fe/H], accounting for each star's measurement uncertainties.
We adopt four components, with the SMC's inner body (the bar and the surrounding main body) described by two overlapping kinematical components, one eastern and one western; a simpler three-component description, with a single inner-body component, remains viable.
For each component we also measure a photometric distance using the tip of the red giant branch (TRGB), which links the chemo-dynamical components to the SMC's line-of-sight structure.

The paper is organized as follows.
Section~\ref{sec:data} describes the data and the sample selection.
Section~\ref{sec:exploratory} presents exploratory kinematical and chemical maps.
Section~\ref{sec:gmm} describes the mixture model and how the number of components is chosen.
Section~\ref{sec:results} presents the properties of the four components, and Section~\ref{sec:discussion} discusses them in the context of the LMC--SMC interaction.
Section~\ref{sec:conclusions} summarizes our conclusions.

\section{Data}
\label{sec:data}

\subsection{SDSS-V Magellanic Genesis spectroscopy}
\label{sec:mgs}

Our spectroscopic sample is drawn from SDSS-V \citep{Kollmeier2026} observations obtained by the Magellanic Genesis Survey \citep[MGS;][]{Nidever2026}, which maps the kinematical and chemical structure of the Magellanic Clouds with wide-field, multi-object spectroscopy.
The spectra were obtained with the Baryon Oscillation Spectroscopic Survey (BOSS) spectrograph \citep{smee2013} on the 2.5~m du Pont Telescope at Las Campanas Observatory \citep{bowen1973}, which SDSS-V operates alongside the 2.5~m Sloan Foundation Telescope at Apache Point Observatory \citep{gunn2006}.
We use the optical BOSS spectra ($R\sim2000$) of red giants that the MGS selected from \gaia\ DR3: stars with proper motions close to the Magellanic systemic motion, with parallaxes consistent with a distant population, and inside a dereddened color--magnitude polygon that extends to $G\approx17.5$ \citep[][their Table~3]{Nidever2026}.
Because of the selection of the parent catalog, the sample contains no stars west of RA~$=0^{\rm h}$. This produces a straight western edge in our sky maps and shapes the spatial distributions of all four components (Section~\ref{sec:overview}).
From these spectra we adopt the stellar parameters, line-of-sight velocities, and metallicities derived from the \textsc{MINESweeper} framework \citep{cargile2020,Chandra2026}.

All velocities and metallicities in our sample therefore come from a single instrument, target selection, and analysis pipeline across the face of the SMC.
Their median per-star uncertainties are 8.1~\kms\ in line-of-sight velocity and 0.13~dex in [Fe/H].

\subsection{\gaia\ DR3 astrometry and photometry}
\label{sec:gaia}

Each star was targeted from \gaia\ DR3 (Section~\ref{sec:mgs}), so we adopt its position, proper motions, and photometry from its \gaia\ DR3 entry, matched by source identifier \citep{gaiadr3}.
Here and throughout, $\mu_{\alpha*}\equiv\mu_\alpha\cos\delta$ denotes the proper motion in right ascension multiplied by $\cos\delta$, and $\mu_\delta$ is the proper motion in declination.
At the magnitudes of our sample, the median proper-motion uncertainty is 0.06~mas~yr$^{-1}$ in each direction, equivalent to $\approx$17~\kms\ at 60~kpc.

Throughout this work we present spatial maps in an SMC-centered Cartesian frame $(X,Y)$, obtained by an orthographic projection of the equatorial coordinates about the adopted SMC center $(\alpha_0,\delta_0)=(13\fdg19,\,-72\fdg83)$:
\begin{align}
X &= \cos\delta\,\sin(\alpha-\alpha_0), \label{eq:projX}\\
Y &= \sin\delta\,\cos\delta_0 - \cos\delta\,\sin\delta_0\,\cos(\alpha-\alpha_0), \label{eq:projY}
\end{align}
with $X$ and $Y$ converted from radians to degrees.
$X$ increases toward the east and $Y$ toward the north; in all sky maps we plot $X$ reversed, so that east is to the left, following the standard on-sky orientation.
Over the extent of our sample, the difference between this projection and a gnomonic (tangent-plane) projection is negligible.

\subsection{Center-of-mass and perspective corrections}
\label{sec:com_frame}

Because the SMC subtends several degrees on the sky, its bulk space motion projects differently onto each line of sight: the systemic motion contributes a position-dependent term to both the observed line-of-sight velocities and the proper motions (the perspective effect; \citealt{vandermarel2002}).
We model this term with the parametric description of the SMC's center-of-mass motion of \citet{zivick2018,zivick2021}, keeping only the systemic-motion and perspective terms.
No internal rotation is subtracted, so any rotation signal remains in the corrected quantities.

For each star, let $v_{\rm los,raw}$ be the heliocentric line-of-sight velocity from \textsc{MINESweeper}, and $v_{\rm los,model}(\alpha,\delta)$ the line-of-sight projection of the SMC's systemic space motion at the star's position.
The center-of-mass-corrected velocity is
\begin{equation}
v_{\rm los,corr} \equiv v_{\rm los,raw} - v_{\rm los,model}(\alpha,\delta).
\label{eq:vcorr}
\end{equation}
Across our field, $v_{\rm los,model}$ varies by about $\pm60$~\kms.
For the proper motions we remove only the perspective term: from each star's proper motion we subtract the variation of the model's systemic proper-motion field across the SMC, which reaches $\approx$0.3~mas~yr$^{-1}$ at the edge of our field.
Unlike $v_{\rm los,corr}$, the corrected proper motions therefore still contain the SMC's bulk motion, which the distance proxy of Section~\ref{sec:dmu} requires.

\subsection{Distance proxy from proper motions}
\label{sec:dmu}

Absolute distances to individual SMC red giants are difficult to measure, but the SMC's shared bulk motion provides a relative distance indicator.
Following the logic of reduced-proper-motion methods \citep[e.g.,][]{Luyten1925}, we define
\begin{equation}
D_{\mu} \equiv \frac{v_{\rm tan}}{4.74\,\mu},
\label{eq:Dmu}
\end{equation}
where $\mu=(\mu_{\alpha*}^2+\mu_\delta^2)^{1/2}$ is the total corrected proper motion in mas~yr$^{-1}$, $v_{\rm tan}$ is the adopted systemic tangential velocity of the SMC in \kms, $D_\mu$ is in kpc, and 4.74~\kms\ is the transverse velocity of an object at 1~kpc moving at 1~mas~yr$^{-1}$.
We adopt $v_{\rm tan}=398$~\kms, as in our previous work \citep{almeida2024}; the median $D_\mu$ of the sample is then 58~kpc.
Because the SMC's systemic transverse velocity ($\approx$400~\kms) is much larger than the internal motions of its stars, a star's total proper motion is very sensitive to its distance, with nearer stars having larger proper motions.
Internal motions, including the tidal expansion modeled by \citet{zivick2021}, also contribute.
We use $D_\mu$ as a relative depth indicator, not as a calibrated distance.

\subsection{Metallicity recalibration with APOGEE}
\label{sec:feh_calib}

\begin{figure}
    \includegraphics[width=\columnwidth]{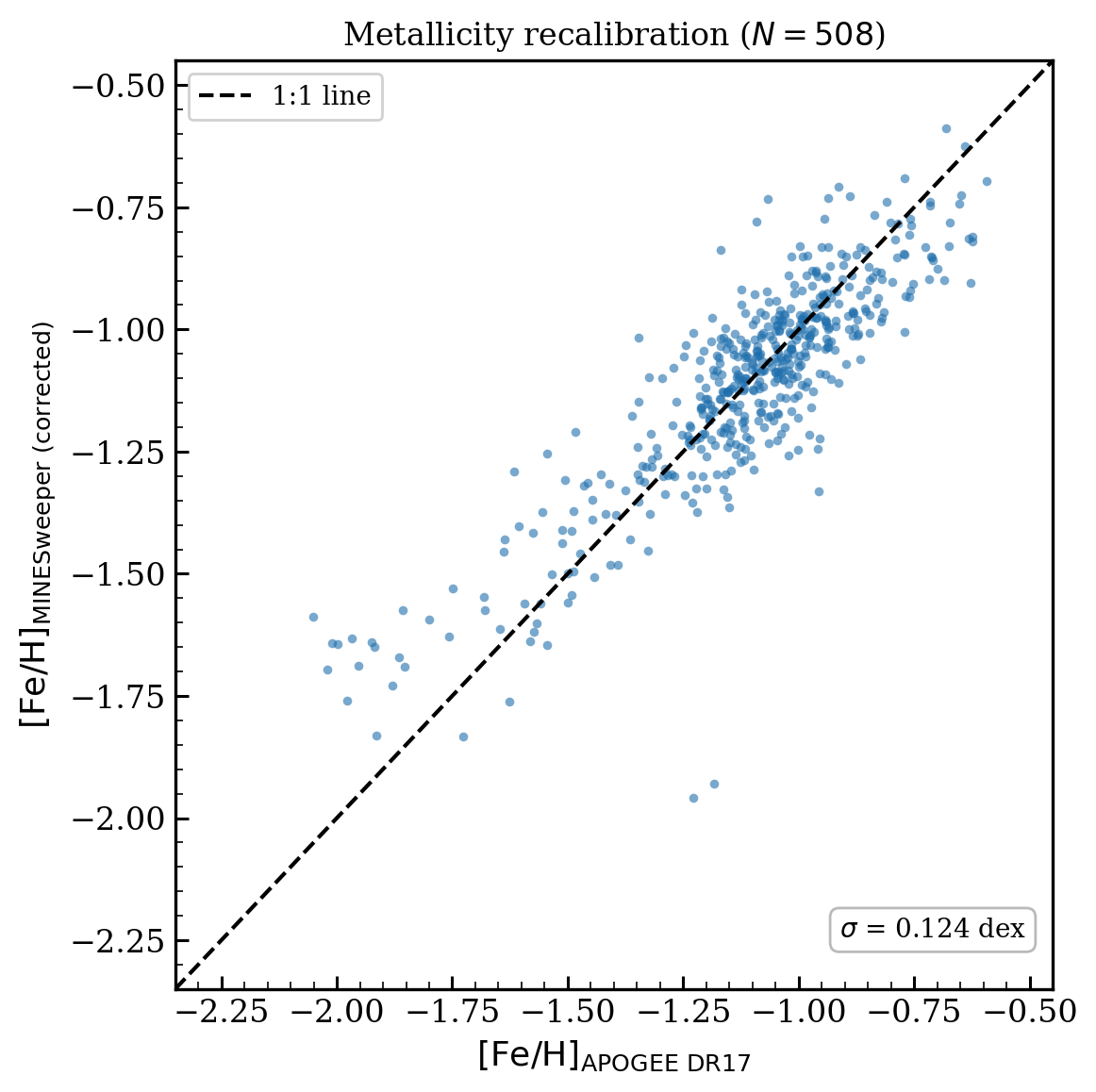}
    \caption{\textsc{MINESweeper} metallicities after the recalibration of Equation~\ref{eq:feh_calib}, [Fe/H]$_{\rm corr}$, against APOGEE DR17 [Fe/H] for the 508 stars in common (S/N\,$>$\,20). The dashed line is the one-to-one relation; the scatter about it is 0.124~dex. The overlap stars are bright upper-RGB stars ($15.6\lesssim G\lesssim17.3$) and span [Fe/H]$_{\rm corr}=-1.96$ to $-0.59$, a range that contains 98\% of our sample. At the metal-poor end the points lie above the line. This is expected from the scatter, because the stars with the lowest APOGEE values include those whose APOGEE values scatter low; grouped by [Fe/H]$_{\rm corr}$ instead, the two scales agree to within 0.08~dex (Section~\ref{sec:feh_calib}).}
    \label{fig:apogee_calib}
\end{figure}

We denote the \textsc{MINESweeper} metallicity by [Fe/H]$_{\rm raw}$.
To place it on a widely used scale, we recalibrate it {\it a posteriori} against APOGEE DR17 [Fe/H] \citep{blanton2017,majewski2017,wilson2019,abdurrouf2022}, using the 508 stars in common with our sample (S/N\,$>$\,20).
These are bright upper-RGB stars ($15.6\lesssim G\lesssim17.3$); the full sample of 8,683 stars extends to the $G\approx17.5$ limit of the MGS selection.
A linear fit of the APOGEE values against [Fe/H]$_{\rm raw}$ gives the recalibrated metallicity
\begin{equation}
[\mathrm{Fe/H}]_{\rm corr} = 0.859\,[\mathrm{Fe/H}]_{\rm raw} - 0.054,
\label{eq:feh_calib}
\end{equation}
with a post-calibration scatter of 0.124~dex (\autoref{fig:apogee_calib}).
The calibration stars span [Fe/H]$_{\rm corr}=-1.96$ to $-0.59$.
At the metal-poor end of \autoref{fig:apogee_calib}, the recalibrated values lie above the one-to-one line, by a median of 0.28~dex for the 14 stars with APOGEE [Fe/H]~$<-1.8$. This is expected from the scatter, because the stars with the lowest APOGEE values include those whose APOGEE values scatter low: grouped instead by [Fe/H]$_{\rm corr}$, the mean APOGEE value agrees with [Fe/H]$_{\rm corr}$ to within 0.08~dex across the whole range, including the 22 calibration stars below $-1.6$, and a quadratic calibration gives essentially the same scatter.
Equivalently, fitting the residual [Fe/H]$_{\rm raw}-$[Fe/H]$_{\rm APOGEE}$ as a linear function of [Fe/H]$_{\rm raw}$ and subtracting it from [Fe/H]$_{\rm raw}$ gives the same transformation.
Metallicities quoted in this paper are on the [Fe/H]$_{\rm corr}$ scale unless stated otherwise.

\subsection{Final sample}
\label{sec:sample}

\begin{figure}
    \includegraphics[width=\columnwidth]{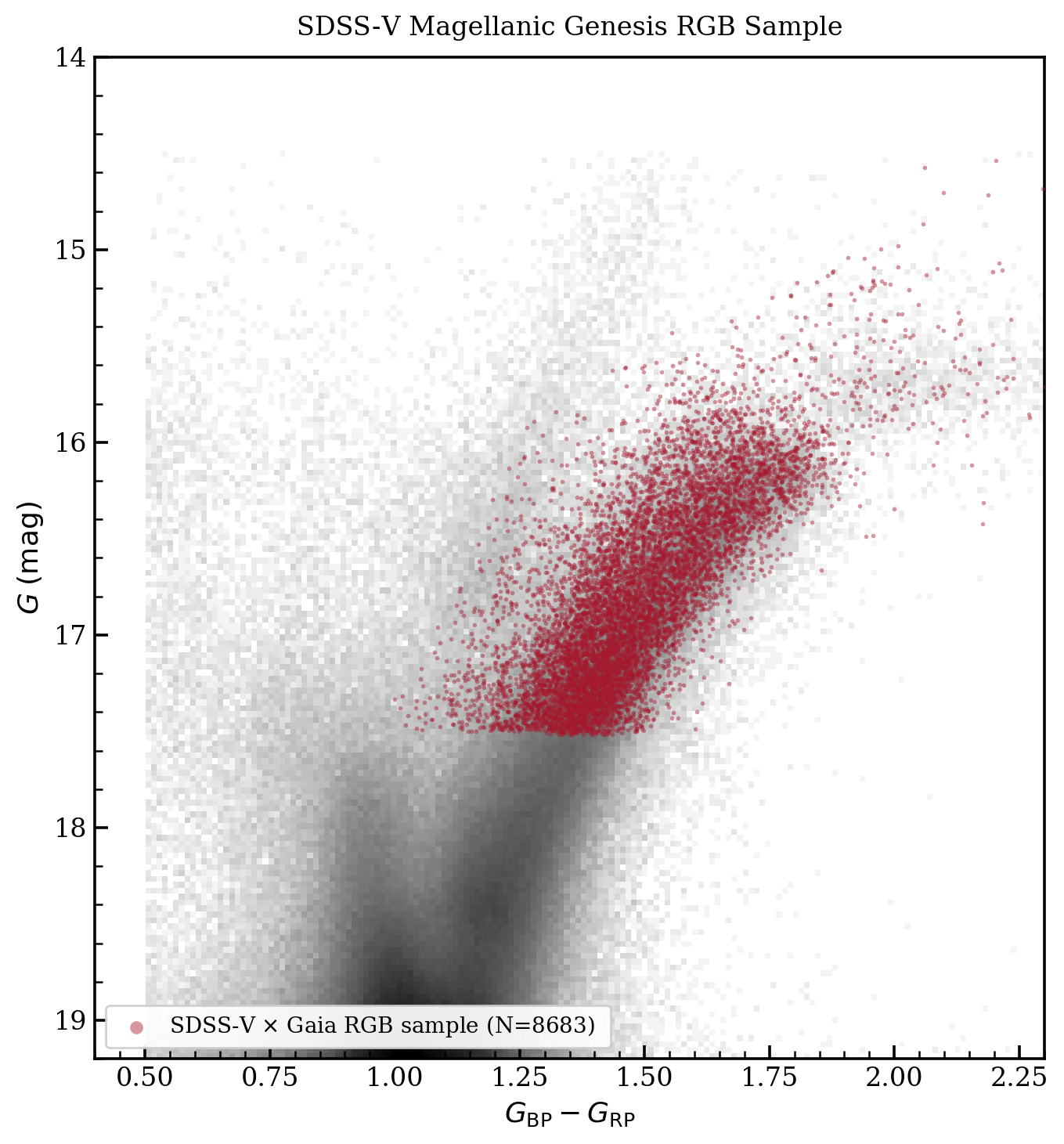}
    \caption{\gaia\ color--magnitude diagram ($G$ versus $G_{\rm BP}-G_{\rm RP}$) of the final sample (red points; $N=8{,}683$), over the density of all \gaia\ sources toward the SMC (grayscale). The sample populates the upper red giant branch, brighter than the $G\approx17.5$ faint limit of the MGS selection (Section~\ref{sec:mgs}).}
    \label{fig:cmd_full}
\end{figure}

For our final sample, we keep giants ($T_{\rm eff}<5{,}500$~K and $\log g<3.5$; a few asymptotic giant branch stars remain, above the tip of the red giant branch in \autoref{fig:cmd_full}) with a well-converged \textsc{MINESweeper} solution and proper motions within the range of those for the SMC ($\mu_{\alpha*}=0.10$ to 1.69 and $\mu_\delta=-1.69$ to $-0.81$~mas~yr$^{-1}$).
SMC membership rests on the MGS proper-motion, parallax, and color--magnitude selection (Section~\ref{sec:mgs}); the line-of-sight velocities of the final sample span 40--260~\kms.

The final sample contains 8,683 stars, each with a line-of-sight velocity, [Fe/H], and \gaia\ proper motions, together with their uncertainties.
\autoref{fig:cmd_full} shows the sample in the \gaia\ color--magnitude diagram: it populates the upper RGB, brighter than the $G\approx17.5$ limit of the MGS selection.

\section{Exploratory analysis}
\label{sec:exploratory}

\begin{figure*}
    \includegraphics[width=\textwidth]{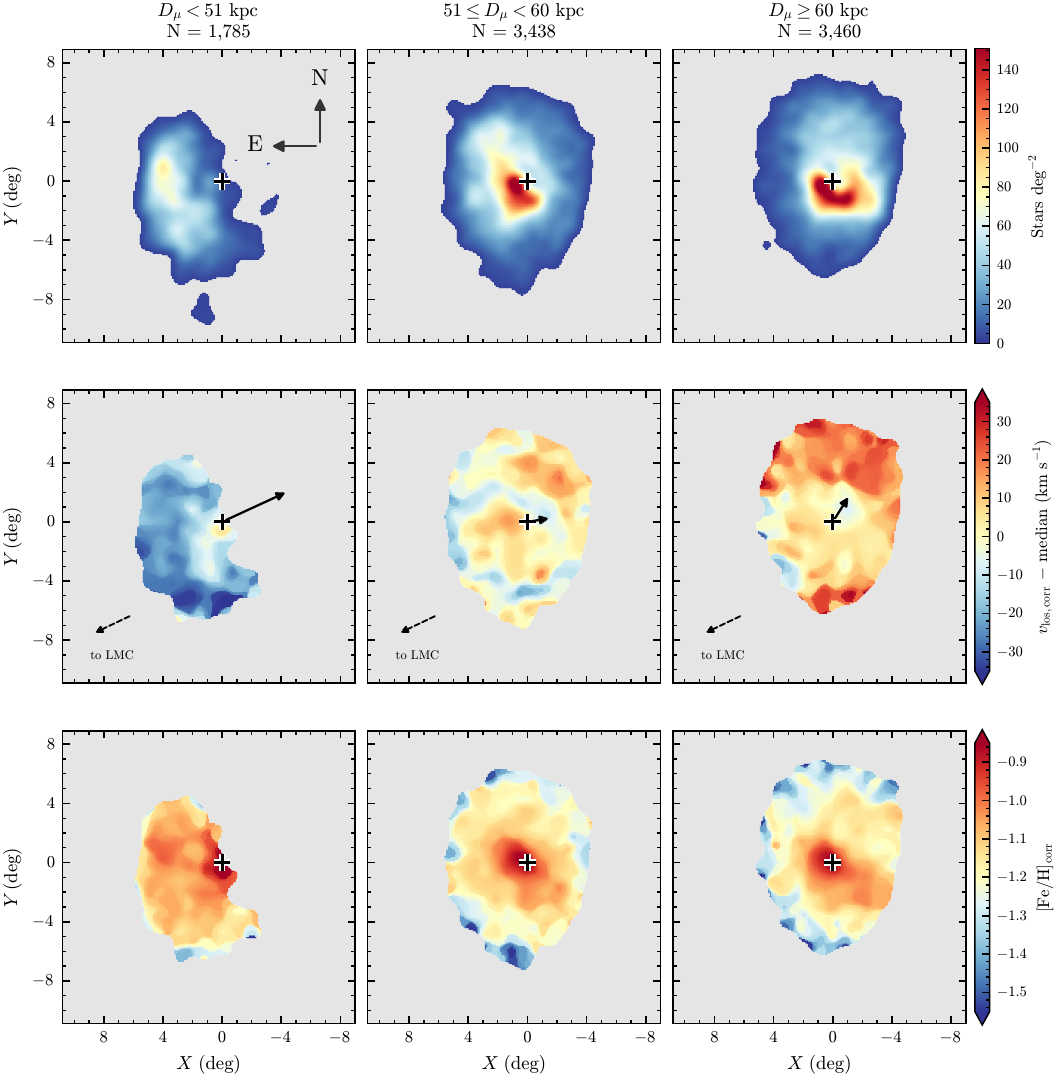}
    \caption{The sample in three slices of the distance proxy $D_\mu$ (Equation~\ref{eq:Dmu}): $D_\mu<51$~kpc (near; left), $51\leq D_\mu<60$~kpc (middle), and $D_\mu\geq60$~kpc (far; right). \textit{Top:} surface density of stars. \textit{Middle:} line-of-sight velocity $v_{\rm los,corr}$ (Equation~\ref{eq:vcorr}) relative to the median of the whole sample. \textit{Bottom:} metallicity [Fe/H]$_{\rm corr}$ (Equation~\ref{eq:feh_calib}). The maps are smoothed with a Gaussian kernel of $\sigma=0.4\degr$ evaluated on 0.1$\degr$ pixels; the velocity and metallicity maps show the kernel-weighted median of the stars within $3\sigma$ of each pixel. A pixel is shown only where the kernel weights of the stars (one at the kernel center) sum to at least five (three in the density maps). Coordinates are the projected $(X,Y)$ frame of Equations~(\ref{eq:projX})--(\ref{eq:projY}), with east to the left; the cross marks the adopted SMC center. Each row shares one color scale. In the middle row, the solid arrow starts at the SMC center and points in the direction in which the velocity increases (Section~\ref{sec:exploratory}), with a length of 1.5$\degr$ per \kms~deg$^{-1}$, and the dashed arrow points toward the LMC.}
    \label{fig:maps_slices}
\end{figure*}

Most of our stars lie within $\approx$6$^{\circ}$ of the SMC center, with some extending to $\approx$8$^{\circ}$ and a few beyond (\autoref{fig:maps_slices}, top row).
Because the SMC has a large line-of-sight depth, a single projected map superposes stars at different distances.
Following the approach of \citet{almeida2024}, we therefore divide the sample into three slices of the distance proxy: $D_\mu<51$~kpc (near; 1,785 stars), $51\leq D_\mu<60$~kpc (middle; 3,438 stars), and $D_\mu\geq60$~kpc (far; 3,460 stars).
Because $D_\mu$ is a relative depth indicator (Section~\ref{sec:dmu}), these slices order the stars coarsely in depth rather than defining physical distance bins.
For each slice, \autoref{fig:maps_slices} shows the surface density of stars, the line-of-sight velocity, and the metallicity.

The three slices have different spatial morphologies.
The near slice is offset to the east: 94\% of its stars lie east of the center, its highest densities are 2--5$\degr$ east of it, and it shows almost no central concentration.
The middle and far slices are dominated by a dense central concentration elongated from northeast to southwest---the SMC bar---embedded in a broader and rounder distribution of stars.

The line-of-sight velocity changes most strongly from one slice to the next.
Relative to the median of the whole sample, the median $v_{\rm los,corr}$ is $-20$, $+2$, and $+8$~\kms\ in the near, middle, and far slices, so the nearer stars approach us relative to the more distant ones.
Within each slice the velocity also varies across the face of the SMC.
A plane fit, $v_{\rm los,corr}=a+bX+cY$, gives gradients of magnitude $(b^2+c^2)^{1/2}=$ $3.3\pm0.4$, $1.2\pm0.2$, and $1.5\pm0.2$~\kms~deg$^{-1}$ in the near, middle, and far slices, with the velocity increasing toward position angles of $295\degr\pm4\degr$, $279\degr\pm9\degr$, and $328\degr\pm10\degr$ (uncertainties from bootstrap resampling of the stars), roughly away from the LMC, which lies at a position angle of $\approx$115$\degr$ as seen from the SMC (arrows in the middle row of \autoref{fig:maps_slices}).
This is the east--west difference in line-of-sight velocity reported by \citet{almeida2024}, with the eastern, LMC-facing side approaching us. \citet{niederhofer2018} map a nonuniform proper-motion field across the central SMC. On the eastern side of the SMC, coherent residual motions point toward the Magellanic Bridge \citep{zivick2018}, along which stars flow from the SMC to the LMC \citep{schmidt2020}.
Here the line-of-sight difference is strongest in the near slice.

Every distance slice has a metal-rich center.
The median [Fe/H]$_{\rm corr}$ of the stars within 1.5$\degr$ of the center is $-0.90$, $-0.95$, and $-0.96$ in the near, middle, and far slices, against $-1.10$, $-1.22$, and $-1.25$ beyond 4$\degr$.
The near slice is comparatively metal-rich throughout, while the two farther slices have more metal-poor outskirts; in these two slices the metal-rich region extends to the southwest, along the SMC bar.

These maps show structures in density, kinematics, and chemistry that overlap both on the sky and along the line of sight.
The near slice is dominated by stars east of the center that approach us relative to the rest of the sample, while in the two farther slices the bar and a broader, more metal-poor distribution occupy the same lines of sight.
Because these populations overlap in every individual observable, no single cut in depth, velocity, or metallicity isolates them.
Separating them requires modeling the joint distribution of line-of-sight velocity, proper motion, and metallicity while accounting for each star's uncertainties, which is the approach we take in Section~\ref{sec:gmm}.

\section{Gaussian mixture modeling}
\label{sec:gmm}

\subsection{The idea}
\label{sec:gmm_idea}

The maps of Section~\ref{sec:exploratory} show that the SMC's stellar populations overlap on the sky, along the line of sight, and in each observable separately.
To separate them, we model the sample as a mixture of a few Gaussian components (``clouds'') in the four-dimensional space of line-of-sight velocity, the two proper-motion components, and metallicity.
The fit finds the center, spread, orientation, and relative size of each cloud in this four-dimensional space, and gives every star the probability that it belongs to each one.
Populations that overlap in any single observable can then be discriminated by how they differ in all four observables together.

The Gaussian form is a practical approximation: each component describes a dominant grouping in the data, which need not be exactly Gaussian or a distinct population (Section~\ref{sec:results}).

Two further choices define the model.
First, each star's measurement uncertainties are accounted for in the model, so that the width of a component reflects the intrinsic spread of its stars rather than the measurement errors, and stars with large uncertainties weigh less in defining the components.
Second, sky position is not used: the components are defined by kinematics and chemistry alone, so their spatial structure is not assumed by the model. The velocities and proper motions we fit are the raw ones, not the corrected ones of Section~\ref{sec:com_frame} (Section~\ref{sec:features}), so they include the position-dependent projection of the SMC's bulk motion, and the mock catalogs of Section~\ref{sec:choosing_k} include it as well.

\subsection{The model}
\label{sec:features}

For each star $i$ we use the four observed quantities
\begin{equation}
\mathbf{x}_i = \left(v_{\rm los,raw},\ \mu_{\alpha*},\ \mu_\delta,\ [\mathrm{Fe/H}]_{\rm raw}\right)_i .
\label{eq:features}
\end{equation}
We fit these raw quantities, rather than the corrected ones of Section~\ref{sec:data}, so that the decomposition does not depend on the adopted model of the SMC's bulk motion; properties measured from the member stars use the corrected quantities.
Before fitting, each quantity is standardized by subtracting its sample mean and dividing by its standard deviation, so that the four enter on comparable scales.
Because of this standardization, the fit is unchanged if a quantity is shifted or rescaled as a whole, with its uncertainties rescaled alike: the linear metallicity calibration of Equation~(\ref{eq:feh_calib}) would give the same decomposition, whereas systematic errors that vary from star to star could alter it.

The model describes the distribution of $\mathbf{x}$ as a sum of $K$ Gaussian components,
\begin{equation}
p(\mathbf{x}) = \sum_{k=1}^{K} \pi_k\, \mathcal{N}(\mathbf{x}\mid\boldsymbol{\mu}_k,\boldsymbol{\Sigma}_k),
\label{eq:mixture}
\end{equation}
where component $k$ has a weight $\pi_k$ (its mixing fraction, the probability that a star drawn from the model belongs to it, with $\sum_k\pi_k=1$; not the number $\pi$), a mean $\boldsymbol{\mu}_k$ (its center in the four quantities), and a covariance matrix $\boldsymbol{\Sigma}_k$ (its spread in each quantity and the correlations between them).
Each component thus has four means, ten independent covariance terms, and a weight, so a model with $K$ components has $p=15K-1$ free parameters.

To include the measurement uncertainties, the model for each star is broadened by that star's error covariance matrix $\mathbf{S}_i$, an approach known as extreme deconvolution \citep{bovy2011}:
\begin{equation}
p(\mathbf{x}_i\mid k) = \mathcal{N}(\mathbf{x}_i\mid\boldsymbol{\mu}_k,\boldsymbol{\Sigma}_k+\mathbf{S}_i).
\label{eq:convolved}
\end{equation}
In other words, component $k$ predicts the observed values of star $i$ by adding the star's measurement scatter to the component's intrinsic spread.
A star with large uncertainties is therefore consistent with several components at once and carries little weight in defining any of them, while $\boldsymbol{\Sigma}_k$ measures the spread intrinsic to the population.
We take $\mathbf{S}_i$ to be diagonal, containing the squared uncertainties of the star's velocity, two proper-motion components, and [Fe/H], each divided by the variance of that quantity in the sample, as for the data.

\subsection{Fitting and memberships}
\label{sec:fitting}

We fit the model with \texttt{pyGMMis} \citep{melchior2018}, which maximizes the likelihood of Equations~(\ref{eq:mixture})--(\ref{eq:convolved}) with the expectation--maximization algorithm.
The algorithm alternates between two steps: given the current components, it computes each star's membership probabilities; given those probabilities, it updates each component's weight, mean, and covariance.
The iterations stop when the log-likelihood changes by less than a fraction $10^{-5}$ between successive steps.
Because the algorithm can settle on a local maximum, we repeated the fit from 15 different starting points; the best of them improves the log-likelihood by only 0.2 and gives almost the same memberships as our fiducial fit (Appendix~\ref{app:modelsel}).

The fit gives, for each star, the probability $P_k$ that it belongs to component $k$, computed with the broadened Gaussians of Equation~(\ref{eq:convolved}).
We assign each star to its most probable component and use these assignments throughout the paper. In the usual terminology for mixture models, this is a hard assignment, and we call the stars assigned to a component its hard members; in a soft assignment, as in the fit itself, each star counts toward every component in proportion to its probabilities $P_k$. We use hard members because the maps, color--magnitude diagrams, and luminosity functions of Section~\ref{sec:results} need a list of stars for each component; the fitted weights, means, and covariances describe the components themselves.
Because the components overlap, hard assignments are imperfect even when the model is exactly right: in a synthetic catalog drawn from our fitted model with the real uncertainties, about one star in four is assigned to the wrong component (Appendix~\ref{app:modelsel}).
The same test shows that the fitted weights, means, and intrinsic dispersions of the components are recovered well.
The properties we measure in Section~\ref{sec:results} come from each component's hard members. Where components overlap, some stars are assigned to the wrong component, so these properties can differ from the fitted parameters; we note where this matters.
As a check, a fit that ignores the measurement uncertainties \citep[with scikit-learn;][]{pedregosa2011} also finds four components and places 87\% of the stars in the same components.

\subsection{How many components?}
\label{sec:choosing_k}

\begin{figure*}
    \includegraphics[width=\textwidth]{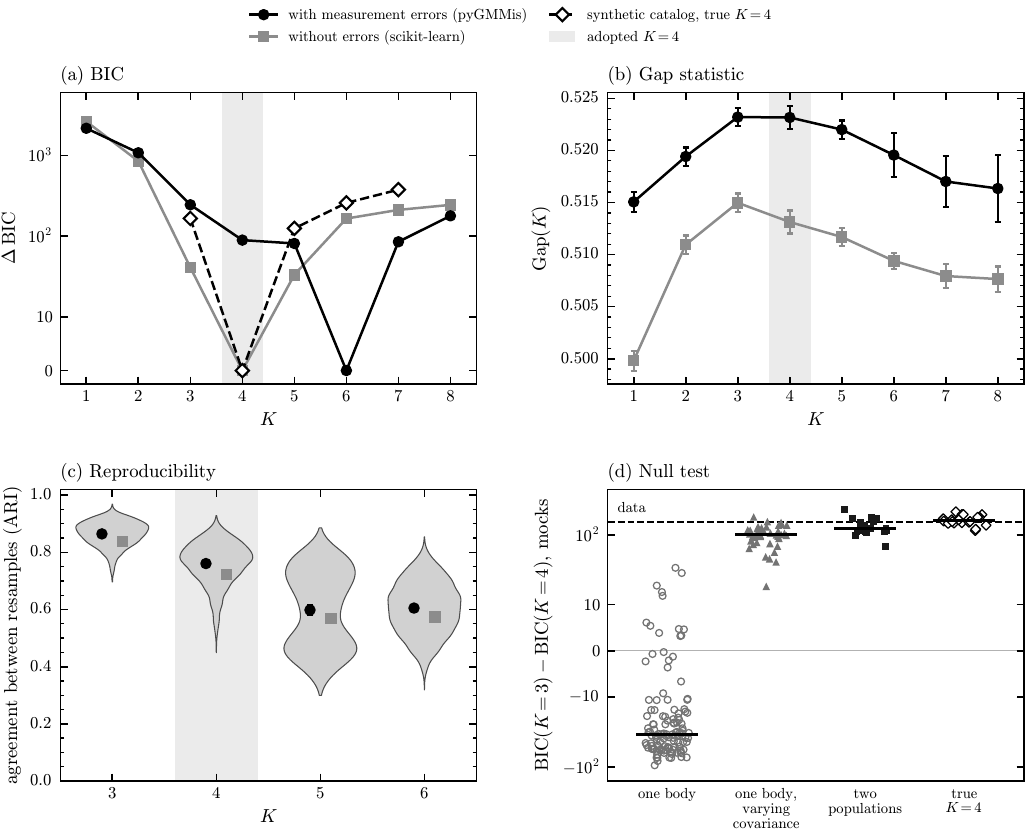}
    \caption{Choosing the number of components $K$ (Section~\ref{sec:choosing_k}). (a) BIC relative to its minimum, for the fit with measurement uncertainties (\texttt{pyGMMis}; black), the fit without them (scikit-learn; gray), and a synthetic catalog drawn from our $K=4$ model with the real uncertainties (open diamonds). (b) Gap statistic and its uncertainty, for the same two fits. (c) Agreement between the memberships of pairs of 100 bootstrap resamples (adjusted Rand index): distribution (violins) and mean (circles) with measurement uncertainties, and mean without them (gray squares). (d) BIC$(K{=}3)-$BIC$(K{=}4)$ for mock catalogs in which the inner body is a single population with smooth gradients (120 mocks), a single population whose intrinsic covariance varies across it (40), or two overlapping populations (20), and in which the whole catalog is drawn from our $K=4$ model (20); horizontal bars mark the medians and the dashed line the data. The vertical axes of (a) and (d) are linear between $-10$ and 10 and logarithmic beyond. Shading marks the adopted $K=4$.}
    \label{fig:modelsel}
\end{figure*}

The number of components, $K$, has to be chosen.
Too few components merge distinct populations; too many split a single population into pieces with no physical meaning.
For our data no single criterion settles the question, so we combine three complementary tests: how well the model fits for its number of parameters, whether its components are reproducible, and whether the split of the SMC's inner body into two components, which appears with four components, could arise from a single body.
\autoref{fig:modelsel} summarizes them, and Appendix~\ref{app:modelsel} gives the details.

The first test balances the quality of the fit against the number of parameters.
The Bayesian information criterion \citep{schwarz1978}, $\mathrm{BIC}=p\ln N-2\ln L_{\rm max}$,
where $L_{\rm max}$ is the maximum likelihood of a model with $p$ parameters fitted to $N$ stars, rewards a better fit (the second term) but penalizes additional parameters (the first term); the preferred model has the lowest BIC.
BIC falls steeply up to $K=4$ and prefers four components to three: BIC$(K{=}3)-$BIC$(K{=}4)=155$ (\autoref{fig:modelsel}a).
Its minimum, however, is at $K=6$.
On a synthetic catalog drawn from our four-component model with the real uncertainties, BIC selects $K=4$, showing that the procedure can recover the input number of components. Its preference for six components in the observed sample suggests that the data depart from four Gaussian components, through non-Gaussian shapes, smaller substructures, or imperfect modeling of the uncertainties. BIC alone cannot establish whether an added Gaussian corresponds to a distinct population; we therefore also examine reproducibility and alternative descriptions of the inner body.
A likelihood-based version of the gap statistic \citep[adapted from][]{tibshirani2001} asks a related question: how much better the model fits the data than it fits structureless reference samples spanning the same range of the four quantities.
The gap is lower for $K\leq2$, indistinguishable for $K=3$ and $K=4$ ($0.5232\pm0.0009$ and $0.5232\pm0.0011$), and declines beyond $K=4$ (\autoref{fig:modelsel}b); its usual selection rule picks the smaller, $K=3$.

The second test asks whether the components are reproducible.
A real component should reappear when the sample is perturbed, whereas an added component that only approximates a non-Gaussian tail in the data need not.
We refit 100 bootstrap resamples of the catalog (stars drawn at random with replacement) and compared the memberships from every pair of resamples with the adjusted Rand index \citep[ARI;][]{hubert1985}, which is 1 for identical memberships and close to 0 for unrelated ones.
The mean agreement is 0.86 for $K=3$ and 0.76 for $K=4$, but drops to 0.60 and 0.61 for $K=5$ and $K=6$ (\autoref{fig:modelsel}c); compared on the same pairs of resamples, $K=4$ is more reproducible than $K=6$ in 94\% of them.
The five- and six-component partitions are therefore much less reproducible.

The first two tests thus leave three or four components: BIC prefers four to three, and reproducibility is higher for three.
The third test asks whether the fourth component, which divides the SMC's inner body into a western and an eastern component (Section~\ref{sec:results} and Appendix~\ref{app:k3}), could instead arise from a single body.
A single inner body whose velocity distribution is not exactly Gaussian could, in principle, also be split in two by the model.
We therefore built mock catalogs in which the 5,798 stars of these two components are replaced by a single population whose mean in each of the four quantities varies smoothly across the sky. Each mock star keeps the position and uncertainties of a real star, the stars of the other two components are kept as observed, and every mock is fitted exactly as the data (Appendix~\ref{app:null}).
In 120 such mocks, BIC$(K{=}3)-$BIC$(K{=}4)$ never exceeds 34, against 155 for the data (\autoref{fig:modelsel}d).
Mocks with two overlapping populations (median 124) and mocks with a single body whose intrinsic spreads and correlations in the four quantities change across it (median 101) both come close to the observed value, and some of them reach it.
The inner body is therefore better described by two Gaussian components than by one with a constant covariance, but the data do not determine whether they are two overlapping populations or one body whose kinematical state varies across it; we refer to them as two kinematical components of the inner body.

Together, these tests support three or four components as a reproducible description of the dominant structure in the sample.
The two descriptions differ only in the inner body, which is one component for $K=3$ and two for $K=4$, while the Foreground Component and the Halo Candidate appear in both (Appendix~\ref{app:k3}).
We adopt $K=4$, which BIC prefers to $K=3$, as our fiducial model, and present the three-component solution in Appendix~\ref{app:k3}.

\section{Properties of the four components}
\label{sec:results}

Section~\ref{sec:gmm} adopted four components as a description of the dominant structures in our sample.
In this section we compare their properties one observable at a time: their distribution on the sky and in the four fitted quantities (Section~\ref{sec:overview}), their color--magnitude diagrams, distances, and ages (Section~\ref{sec:distances}), their kinematics (Section~\ref{sec:kinematics}), their chemistry (Section~\ref{sec:chemistry}), and their maps across the face of the SMC (Section~\ref{sec:comp_maps}).
Throughout, we use hard members and the corrected quantities of Section~\ref{sec:data}: line-of-sight velocities are given relative to the median $v_{\rm los,corr}$ of the whole sample, as in Section~\ref{sec:exploratory}, and proper motions are perspective-corrected.
We interpret the components in Section~\ref{sec:discussion}.

\subsection{The components on the sky and in the four observables}
\label{sec:overview}

We name the components after properties described in this section: the Western Inner Body (3,063 stars), the Foreground Component (1,638), the Halo Candidate (1,247), and the Eastern Inner Body (2,735).
Their hard members make up 35\%, 19\%, 14\%, and 32\% of the sample.
In the three-component model, the two inner bodies merge into one component, while the Foreground Component and the Halo Candidate keep their identity (Appendix~\ref{app:k3}).

The components are defined by line-of-sight velocity, proper motion, and metallicity alone.
Three of the properties we examine below do not enter the fit: the color--magnitude diagrams and the distances from the tip of the red giant branch (Section~\ref{sec:distances}), and the APOGEE abundances (Section~\ref{sec:chemistry}).
They test whether the components differ as stellar populations, and not only in the quantities used to define them.

\begin{figure*}
    \includegraphics[width=\textwidth]{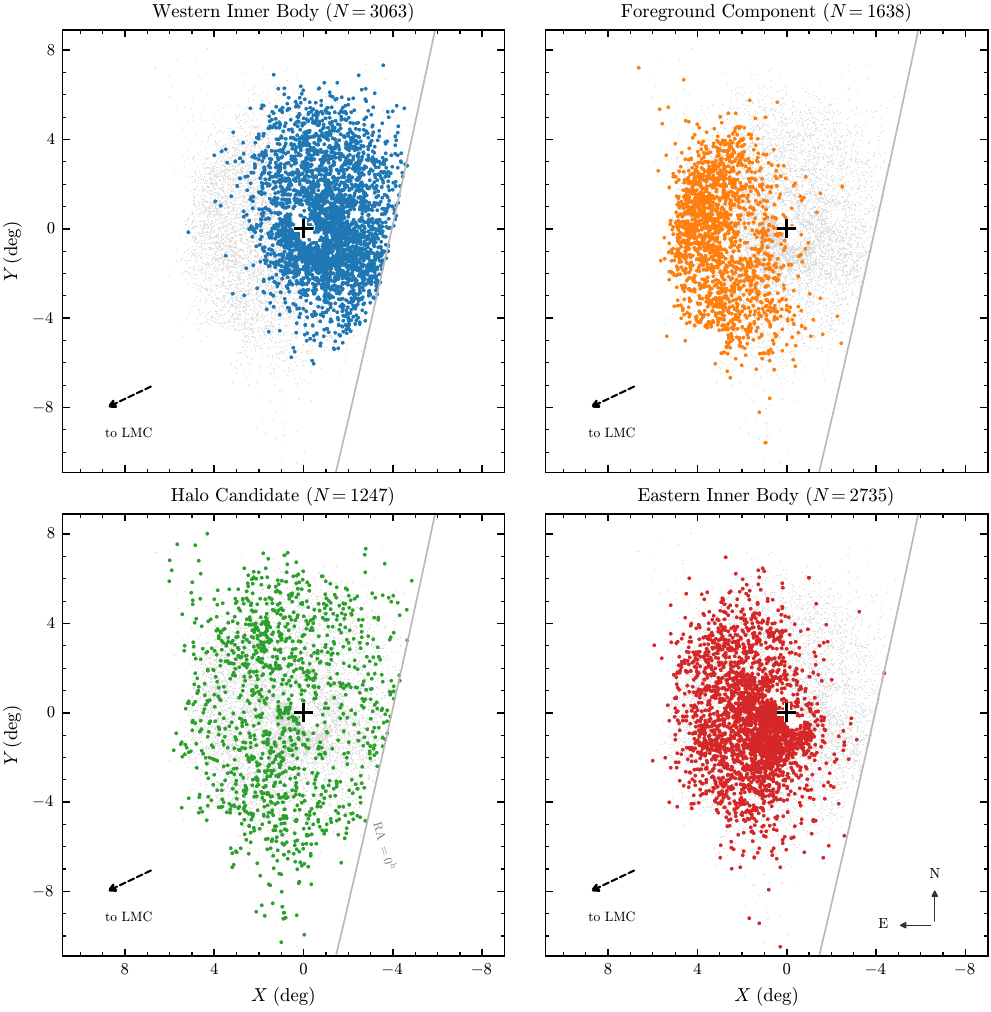}
    \caption{The four components on the sky: hard members of each component (colors) over the whole sample (gray), in the projected $(X,Y)$ frame of Equations~(\ref{eq:projX})--(\ref{eq:projY}), with east to the left. The cross marks the adopted SMC center; panel titles give the number of members. The dashed arrow points toward the LMC. The light gray line marks RA~$=0^{\rm h}$, the western limit of the sample (Section~\ref{sec:mgs}).}
    \label{fig:sky}
\end{figure*}

\autoref{fig:sky} shows the components on the sky.
The two inner bodies overlap at the center of the SMC but are displaced from each other: the centroid (median position) of the Western Inner Body lies 1.0$\degr$ west of the center, and that of the Eastern Inner Body 1.4$\degr$ east.
The Foreground Component is the most offset, with its centroid 3.1$\degr$ east of the center and 97\% of its stars on the eastern side.
The Halo Candidate is the most extended (the median distance of its stars from its centroid is 3.9$\degr$, versus 2.1--2.4$\degr$ for the other three), and its centroid, 1.1$\degr$ east of the center, is close to that of the whole sample (0.9$\degr$), as expected for a component spread over the whole footprint.
The survey footprint itself, which is elongated north--south on the sky and ends at RA~$=0^{\rm h}$ (Section~\ref{sec:mgs}), shapes all four distributions, so we compare the components through their centroids and sizes within the footprint rather than through their shapes.

\begin{figure*}
    \includegraphics[width=\textwidth]{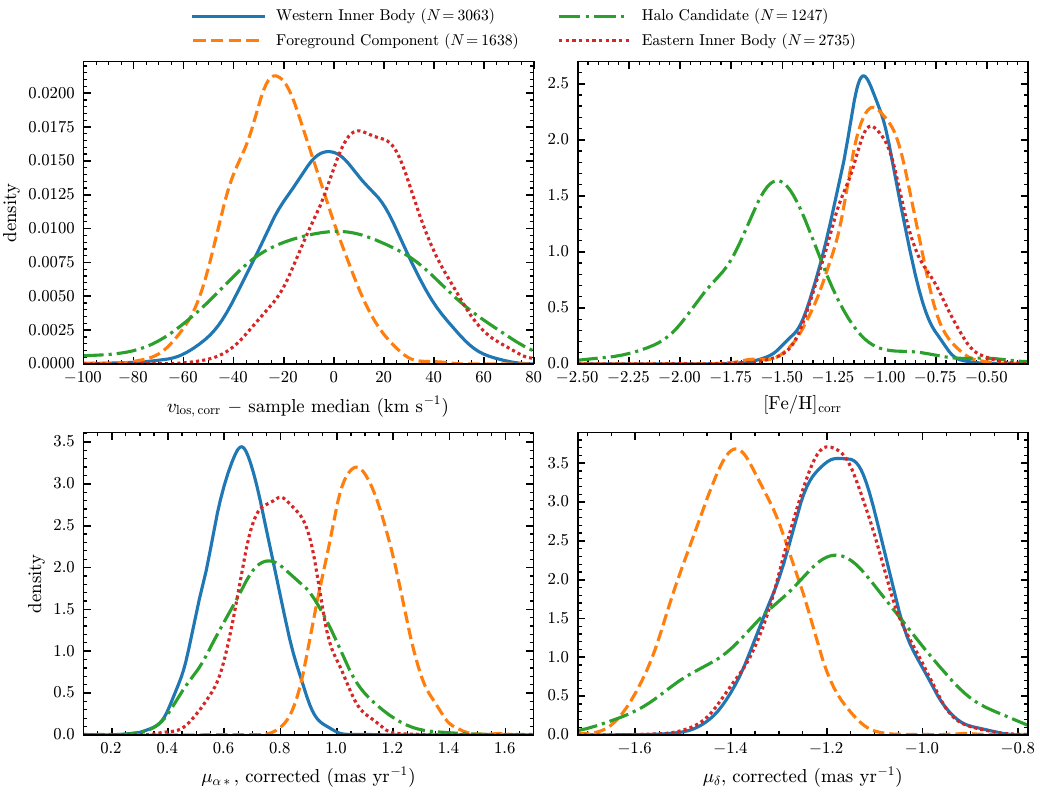}
    \caption{Distributions of the four fitted quantities for the four components (hard members; kernel density estimates, each normalized to unit area): line-of-sight velocity relative to the median of the whole sample, metallicity, and the two perspective-corrected proper-motion components. The mixture is fitted to the raw quantities (Section~\ref{sec:features}); the corrected ones are shown so that the modeled bulk motion and perspective, which vary with position, do not broaden the distributions. Line styles distinguish the components independently of color.}
    \label{fig:dist4}
\end{figure*}

\autoref{fig:dist4} shows the distributions of the four fitted quantities.
No single quantity separates all four components, but each component stands out in at least one.
In line-of-sight velocity, the median of the Foreground Component lies 22~\kms\ below that of the whole sample and that of the Eastern Inner Body 14~\kms\ above it, while the Western Inner Body and the Halo Candidate lie within 1~\kms\ of it; the Halo Candidate has the broadest distribution.
In metallicity, the Halo Candidate is the most metal-poor component (median [Fe/H]$_{\rm corr}=-1.54$, with 16th--84th percentiles $-1.83$ and $-1.31$), while the other three are similar ($-1.09$, $-1.04$, and $-1.05$ for the Western Inner Body, the Foreground Component, and the Eastern Inner Body).
In proper motion, the Foreground Component is offset from the other three in both directions (median $\mu_{\alpha*}=1.09$ and $\mu_\delta=-1.39$~mas~yr$^{-1}$, against 0.77 and $-1.22$~mas~yr$^{-1}$ for the whole sample), which is the clearest separation in the four quantities; the Western Inner Body has a lower $\mu_{\alpha*}$ (0.66~mas~yr$^{-1}$) than the Eastern Inner Body (0.80~mas~yr$^{-1}$), and the Halo Candidate again has the broadest distributions.
We describe the kinematics of the components in Section~\ref{sec:kinematics} and their chemistry in Section~\ref{sec:chemistry}.

Because the MGS selects stars by proper motion, parallax, and color (Section~\ref{sec:mgs}), Milky Way giants with SMC-like proper motions can remain in the sample.
\citet{Nidever2026} estimate a contamination of about 1\% within 6$\degr$ of the SMC and 5--10\% beyond, or 97--110 stars in our sample.
Such stars would most likely be assigned to the Halo Candidate, the broadest component.
Its 133 members beyond 6$\degr$ (11\% of the component) have the same line-of-sight velocity dispersion and median metallicity as its members within 4$\degr$ (37 and 39~\kms; [Fe/H]$_{\rm corr}=-1.53$ in both), so we see no sign of a distinct contaminating population in its outskirts.
If every expected contaminant were assigned to it, they would make up about 8\%--9\% of the Halo Candidate.

\subsection{Color--magnitude diagrams, distances, and ages}
\label{sec:distances}

\begin{figure*}
    \includegraphics[width=\textwidth]{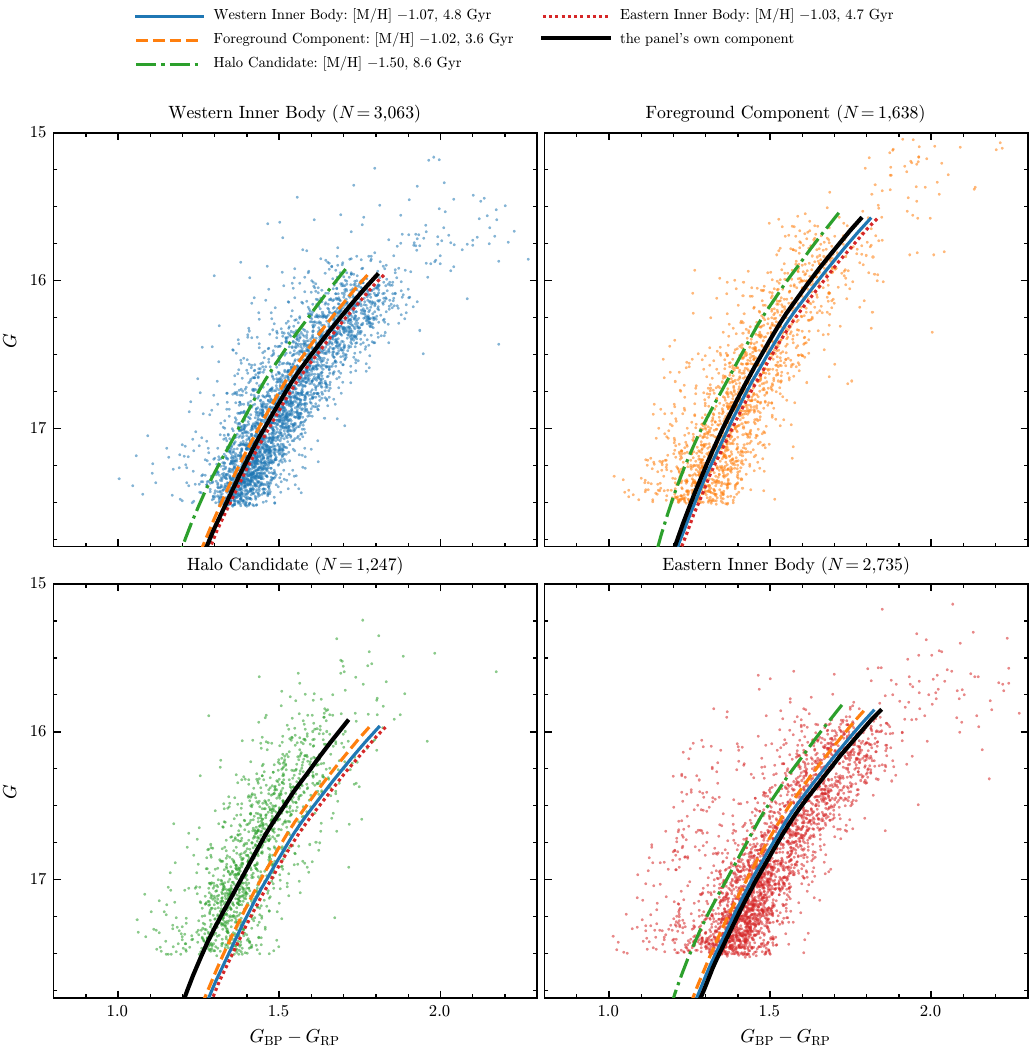}
    \caption{Color--magnitude diagrams of the four components (hard members). Every panel shows the PARSEC isochrones of all four components (Section~\ref{sec:distances}), placed at that panel's distance and at the median extinction of its members, so that the curves differ only in metallicity and age; the panel's own isochrone is solid black, and the others are drawn in their component's color and line style, as in \autoref{fig:dist4}. Distances are from the TRGB, except for the Halo Candidate, which is placed at the TRGB distance of the SMC as a whole. The isochrones are interpolated in metallicity and age between those of our grid (for the two inner bodies, extrapolated in age beyond 4~Gyr) and are shown up to the tip of the red giant branch; most stars above it are asymptotic giant branch stars. The key gives the adopted metallicity and age of each component; for the Halo Candidate, changes in the adopted distance, extinction, and metallicity give ages between 6.4 and 11.3~Gyr (Section~\ref{sec:distances}). The stars bluer than the red giant branch, most visible in the Eastern Inner Body, are younger SMC stars admitted by the color selection of the survey (Section~\ref{sec:mgs}).}
    \label{fig:cmd}
\end{figure*}

\autoref{fig:cmd} shows the color--magnitude diagrams of the four components.
Each component populates a red giant branch (RGB) that ends at a well-defined tip, with a few asymptotic giant branch (AGB) stars above it.
The RGB of the Foreground Component is brighter than those of the other three, by 0.3--0.4~mag at the tip, and the RGB of the Halo Candidate is bluer than the others.
We use the tip of the RGB (TRGB) to measure the distance of each component independently of its kinematics, and the color of the RGB to compare the ages of the components.
The process for determining the TRGB for each population is as follows.

We correct for extinction with a map of \textsc{MINESweeper} extinctions: the median $A_V$ of all the stars in 0.75$\degr$ cells.
\textsc{MINESweeper} fits the extinction of each star from its spectrum and photometry, together with its other parameters \citep{cargile2020}.
For comparison, the dust map of \citet{schlegel1998}, recalibrated by \citet{schlafly2011}, gives a lower extinction at the Halo Candidate's members of $A_V=0.11$, versus 0.14 from our map; this difference sets the extinction variation used for the age limits discussed below.
For the isochrone fits, we deredden each star with the value of our map at its position.
For the TRGB, we use the median of these values over each component's members, $A_V=0.12$, 0.14, 0.14, and 0.17 for the Western Inner Body, the Foreground Component, the Halo Candidate, and the Eastern Inner Body.
We convert $A_V$ to the \gaia\ bands with the coefficients of \citet{wang2019}: $A_G=0.789\,A_V$, $A_{\rm RP}=0.589\,A_V$, and $E(G_{\rm BP}-G_{\rm RP})=0.413\,A_V$.

\begin{figure*}
    \includegraphics[width=\textwidth]{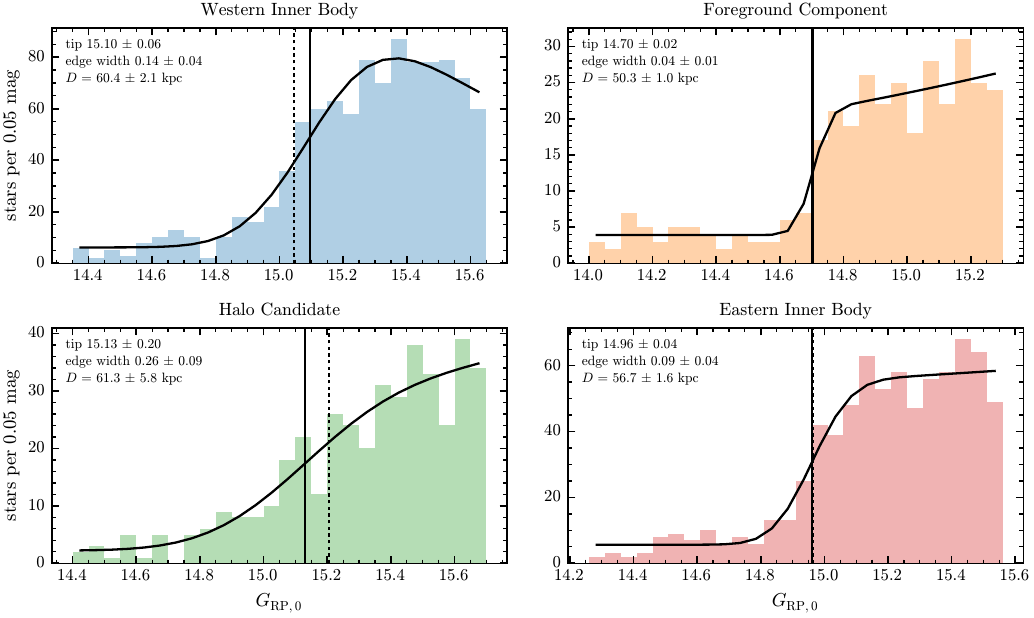}
    \caption{Luminosity functions of the four components near the tip of the red giant branch, in dereddened \gaia\ $G_{\rm RP}$ magnitudes (histograms, 0.05~mag bins; only one of the five binnings is shown). Black curves show the model fitted to that binning: an exponential luminosity function multiplied by a smooth step centered on the tip (a Gaussian cumulative distribution), plus a constant level for the brighter asymptotic giant branch stars. Solid vertical lines mark the tip averaged over the five binnings, and dotted lines the tip from a Sobel edge filter. Each panel gives the tip, the edge width, and the resulting distance (Section~\ref{sec:distances}).}
    \label{fig:trgb}
\end{figure*}

The tip is usually located with a Sobel edge filter \citep{lee1993}, which marks the magnitude at which the smoothed luminosity function changes most steeply. We fit a model of the luminosity function instead, because it gives the width of the edge as well as the tip, and we compare the two estimates below.
To determine the precise location of the TRGB for each component, we fit the luminosity function of its hard members in dereddened \gaia\ $G_{\rm RP}$ (\autoref{fig:trgb}).
The model is an exponential RGB luminosity function that is cut off at its bright end by a smooth step (a Gaussian cumulative distribution), plus a constant level for the brighter AGB stars. The tip is the magnitude at the center of this step, and the Gaussian width of the step is the width of the edge.
We fit it by maximizing a Poisson likelihood in 0.05~mag bins, evaluating the model at the bin centers.
The tip fitted to a single binning shifts with the positions of the bin edges, by up to 0.08~mag for the Western Inner Body and 0.24~mag for the Halo Candidate, so we average the fits over five binnings offset from one another by 0.01~mag; uncertainties come from 150 bootstrap resamples, each fitted in the same way.
We convert the tip to a distance with the \gaia-based calibration $M_{\rm RP,0}^{\rm TRGB}=-3.807^{+0.041}_{-0.035}$ \citep{cruzreyes2026}, adding its uncertainty in quadrature.
Because the calibration locates the tip with this filter, on the luminosity function smoothed with a 0.125~mag kernel \citep{cruzreyes2026}, we also applied it to our luminosity functions: it gives tips within 0.05~mag of the fitted ones for the two inner bodies and the Foreground Component, and 0.08~mag fainter for the Halo Candidate, less than the uncertainties of the tips.

The TRGB distances are $60.4\pm2.1$, $56.7\pm1.6$, $61.3\pm5.8$, and $50.3\pm1.0$~kpc for the Western Inner Body, the Eastern Inner Body, the Halo Candidate, and the Foreground Component.
The Foreground Component lies $6.3\pm1.2$~kpc in front of the Eastern Inner Body and $10.0\pm1.8$~kpc in front of the Western Inner Body, and the Eastern Inner Body is $3.7\pm2.1$~kpc (1.8$\sigma$) nearer than the Western Inner Body.
The uncertainties of these differences do not include that of the calibration, which shifts all the distances together.
The distance of the Halo Candidate is the least certain; it agrees with those of the inner bodies and with the TRGB distance of the SMC as a whole, 60.2~kpc \citep[$\mu=18.90$;][]{cruzreyes2026}.
We adopt the TRGB distances in the rest of this paper, to convert proper motions to velocities and to place the isochrones.

\begin{figure}
    \includegraphics[width=\columnwidth]{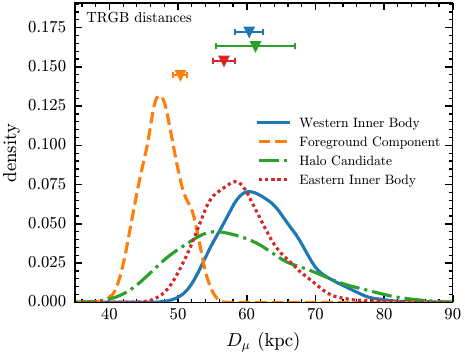}
    \caption{Distribution of the distance proxy $D_\mu$ (Equation~\ref{eq:Dmu}) for the four components (hard members; kernel density estimates, each normalized to unit area). Triangles mark the TRGB distances, with error bars showing their uncertainties (Section~\ref{sec:distances}).}
    \label{fig:dmu}
\end{figure}

The distance proxy $D_\mu$ agrees broadly with the TRGB (\autoref{fig:dmu}): it also places the Foreground Component nearest (median 47.6~kpc) and the Western Inner Body farther than the Eastern Inner Body (61.8 and 58.4~kpc), with the Halo Candidate at 58.0~kpc.
These medians agree with the TRGB distances to within 0.12~mag.
Because $D_\mu$ is computed from the proper motions (Equation~\ref{eq:Dmu}), which are fitted quantities, its ordering of the components is not an independent distance measurement; the agreement with the TRGB, which does not use the proper motions, is what supports $D_\mu$ as the relative depth indicator used for the slices of Section~\ref{sec:exploratory}.
The widths of the $D_\mu$ distributions (the median absolute deviation scaled by 1.4826, so that it estimates the standard deviation for a Gaussian distribution; 7\%--15\% of the median) include the proper-motion errors (4\% of the bulk proper motion) and the internal motions of the stars along the direction of the bulk motion, so they should not be read as depths along the line of sight.

The width of the edge also limits the spread in distance within each component.
The Foreground Component has the sharpest edge, $0.04\pm0.01$~mag; even if all of it were due to depth, its depth dispersion would be below 1.2~kpc (1$\sigma$), so this component is a sheet that is thin along the line of sight.
If the extra width of each of the other edges, beyond the Foreground Component's, comes from a spread in distance ($\sigma_D=0.4605\,D\,\sigma_{\rm edge}$, with the two widths subtracted in quadrature), the depth dispersions are 3.9, 2.2, and 7.2~kpc for the Western Inner Body, the Eastern Inner Body, and the Halo Candidate (at most 5.1, 3.4, and 9.9~kpc, 1$\sigma$).
These are upper limits, since spreads in age and metallicity and contamination by other components also broaden the edge.

The colors of the RGBs order the components in age.
We fit PARSEC isochrones \citep{bressan2012,marigo2017} in the \gaia\ EDR3 bands to the RGB of each component, at its distance and extinction and at its median metallicity: for each metallicity of our isochrone grid, we find the age at which the isochrone runs through the median color of the RGB in 0.25~mag bins, and we interpolate these ages to the metallicity of the component.
Because the isochrones have scaled-solar abundances, we use $[{\rm M/H}]=[{\rm Fe/H}]+\log(0.638\,f_\alpha+0.362)$ \citep{salaris1993}, with $f_\alpha=10^{[\alpha/{\rm M}]}$ from the median APOGEE abundances of each component (Section~\ref{sec:chemistry}); this correction is 0.02--0.04~dex.
For the Halo Candidate, whose own tip is the least certain, we adopt the TRGB distance of the SMC as a whole.
The Western Inner Body, the Eastern Inner Body, and the Foreground Component are then 4.8, 4.7, and 3.6~Gyr old (for the first two, extrapolated slightly beyond 4~Gyr, the oldest isochrone of our grid at their metallicity), and the Halo Candidate is 8.6~Gyr old: its RGB is 0.07--0.10~mag bluer than the isochrones of the other three components (\autoref{fig:cmd}).
These absolute ages depend on the isochrones: MIST isochrones \citep{choi2016} give older ages for all four components, while also making the Halo Candidate the oldest. They also depend on the adopted distance, extinction, and metallicity: changes of 0.05~mag in distance modulus, 0.03~mag in $A_V$, and 0.1~dex in [M/H] change the age of the Halo Candidate by 1--3~Gyr, to between 6.4 and 11.3~Gyr.
We use them only to note that the Halo Candidate is the oldest component, about 1.8 times older than the inner bodies.

\subsection{Kinematics}
\label{sec:kinematics}

We describe the motions of the components along the line of sight and, in the plane of the sky, along and across the direction of the SMC's bulk proper motion (position angle 148$\degr$), which we call the bulk-motion axis; it is a direction on the sky, not the line of sight.
Because the SMC's bulk motion lies along this axis, a star's proper motion along it is larger if the star is nearer (the dependence behind $D_\mu$; Section~\ref{sec:dmu}) or if it moves faster along the axis; proper motions alone cannot separate these effects. A difference in distance changes the bulk-motion contribution along this axis, but does not create one across it.
We convert proper motions to velocities at each component's TRGB distance (Section~\ref{sec:distances}).
\autoref{tab:kin} collects the measurements.

\begin{deluxetable*}{lcccc}
\tablecaption{Properties of the four components \label{tab:kin}}
\tablehead{
\colhead{Quantity} & \colhead{Western Inner Body} & \colhead{Foreground Component} & \colhead{Halo Candidate} & \colhead{Eastern Inner Body}
}
\startdata
\sidehead{Membership and metallicity}
$N$ (hard members) & 3,063 & 1,638 & 1,247 & 2,735 \\
Hard members (\% of the sample) & 35 & 19 & 14 & 32 \\
Median [Fe/H]$_{\rm corr}$ & $-1.09$ & $-1.04$ & $-1.54$ & $-1.05$ \\
16th and 84th percentiles of [Fe/H]$_{\rm corr}$ & $-1.25$, $-0.93$ & $-1.21$, $-0.88$ & $-1.83$, $-1.31$ & $-1.23$, $-0.84$ \\
\sidehead{Position and distance}
Centroid $X$ (deg) & $-1.0$ & $+3.1$ & $+1.1$ & $+1.4$ \\
Centroid $Y$ (deg) & $+0.1$ & $-0.2$ & $+0.5$ & $-0.5$ \\
Median distance from the centroid (deg) & 2.3 & 2.4 & 3.9 & 2.1 \\
TRGB distance (kpc) & $60.4\pm2.1$ & $50.3\pm1.0$ & $61.3\pm5.8$ & $56.7\pm1.6$ \\
Line-of-sight depth dispersion (kpc) & 3.9 ($<$5.1) & $<$1.2 & 7.2 ($<$9.9) & 2.2 ($<$3.4) \\
\sidehead{Kinematics}
$\sigma_{\rm los}$ (\kms) & $24.3\pm0.6$ & $17.3\pm0.6$ & $37.9\pm1.2$ & $20.8\pm0.5$ \\
$\sigma_\perp$ (\kms) & $21.9\pm0.5$ & $22.6\pm0.7$ & $39.2\pm1.2$ & $25.1\pm0.6$ \\
$\sigma_\parallel$ (\kms) & $30.4\pm0.5$ & $22.0\pm0.5$ & $59.3\pm1.1$ & $29.8\pm0.5$ \\
Axis ratio of the plane-of-sky velocity ellipse & $0.64\pm0.02$ & $0.86\pm0.04$ & $0.61\pm0.02$ & $0.67\pm0.02$ \\
Position angle of its major axis (deg) & $122\pm2$ & $143\pm9$ & $136\pm2$ & $123\pm2$ \\
$\Delta V_{\rm los}$ (\kms) & $-3.7\pm2.1$ & $-12.3\pm2.1$ & \nodata & $+19.9\pm2.1$ \\
$\Delta v_\perp$ (\kms) & $+27.6\pm2.2$ & $-30.4\pm2.6$ & \nodata & $-6.3\pm2.2$ \\
$\Delta v_\parallel$ (\kms) & $-18\pm42$ & $-18\pm41$ & \nodata & $-41\pm41$ \\
$(\Delta V_{\rm los}^2+\Delta v_\perp^2)^{1/2}$ (\kms) & $27.9$ & $32.8$ & \nodata & $20.9$ \\
$|\nabla v_{\rm los}|$ (\kms~deg$^{-1}$) & $5.1\pm0.2$ & $3.3\pm0.3$ & $3.4\pm0.5$ & $5.3\pm0.3$ \\
Position angle of $\nabla v_{\rm los}$ (deg) & $301\pm2$ & $314\pm4$ & $271\pm5$ & $296\pm2$ \\
\enddata
\tablecomments{Hard members. [Fe/H]$_{\rm corr}$ is the recalibrated metallicity of Equation~(\ref{eq:feh_calib}). Centroids are median positions in the $(X,Y)$ frame of Equations~(\ref{eq:projX})--(\ref{eq:projY}), with $X$ increasing toward the east and $Y$ toward the north, and sizes are median angular distances from the centroid. Depth dispersions come from the widths of the TRGB edges (Section~\ref{sec:distances}): for the Foreground Component, the 1$\sigma$ upper limit if its whole edge width were due to depth; for the others, the width in excess of the Foreground Component's, with the 1$\sigma$ upper limit in parentheses. Velocity dispersions are intrinsic (measurement errors removed, as in the fit); $\sigma_{\rm los}$ is the scaled median absolute deviation of $v_{\rm los,corr}$. Plane-of-sky velocities are computed at each component's TRGB distance; $\perp$ and $\parallel$ denote the directions across and along the bulk-motion axis (position angle 148$\degr$), with positive $\perp$ toward position angle 238$\degr$. $\Delta$ quantities are motions relative to the Halo Candidate, measured against its stars in the same 1.5$\degr$ cells; the uncertainty of $\Delta v_\parallel$ includes those of the two TRGB distances. Gradients are plane fits to $v_{\rm los,corr}$, with the position angle toward which the velocity increases. Uncertainties are bootstrap standard deviations (for the TRGB distances, the scaled median absolute deviation of the bootstrap tips, with the 0.04~mag calibration uncertainty added). Except for $\Delta v_\parallel$, the uncertainties of the plane-of-sky velocities are at fixed distance; these velocities scale with the adopted distances, whose uncertainties are 2\%--9\%.}
\end{deluxetable*}

\subsubsection{Velocity dispersions}

\begin{figure*}
    \includegraphics[width=\textwidth]{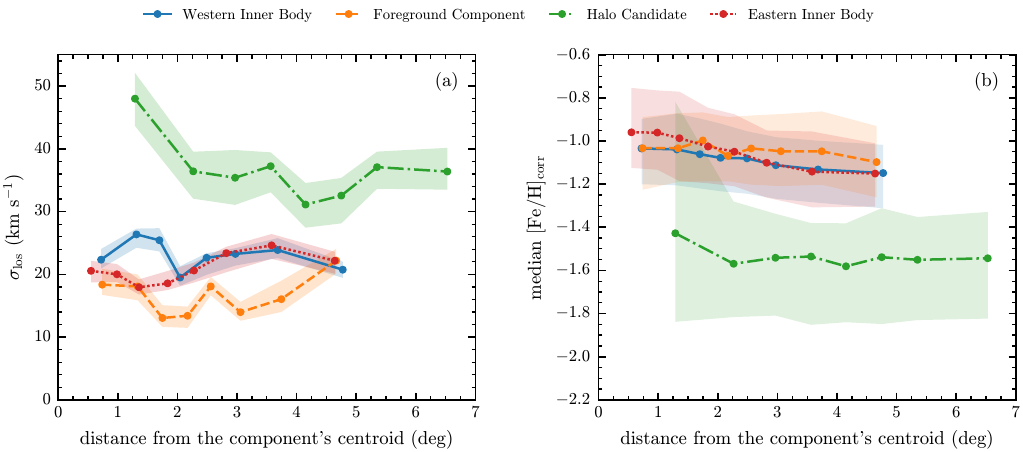}
    \caption{Radial profiles of the four components, as a function of the distance of each star from its component's centroid (eight bins with equal numbers of stars per component). (a) Line-of-sight velocity dispersion, the scaled median absolute deviation of $v_{\rm los,corr}$ with the median measurement error removed in quadrature; bands show the 16th--84th percentiles of 400 bootstrap resamples. (b) Median [Fe/H]$_{\rm corr}$; bands show the 16th--84th percentiles of the stars in each bin.}
    \label{fig:radial}
\end{figure*}

The Halo Candidate is the hottest component along the line of sight: $\sigma_{\rm los}=37.9\pm1.2$~\kms, against $24.3\pm0.6$, $20.8\pm0.5$, and $17.3\pm0.6$~\kms\ for the Western Inner Body, the Eastern Inner Body, and the Foreground Component.
These are the scaled median absolute deviations of $v_{\rm los,corr}$, with the median measurement error removed in quadrature.
The Halo Candidate has the largest dispersion at every sampled distance from each component's own centroid (\autoref{fig:radial}a). The two inner bodies differ most between 1$\degr$ and 2$\degr$ from their centroids, where the Western Inner Body is the hotter; beyond 2$\degr$ their dispersions agree within the bootstrap bands.
Near its centroid, the Halo Candidate's dispersion rises to $\approx$48~\kms. There the inner bodies dominate, so stars with velocities close to theirs are preferentially assigned to the inner bodies, and the Halo Candidate's hard members are mostly stars in the wings of the velocity distribution; the rise therefore probably reflects the hard assignment rather than an intrinsically hotter central region (Section~\ref{sec:fitting}).

In the plane of the sky, the dispersions across the bulk-motion axis are 21.9, 25.1, 22.6, and 39.2~\kms\ for the Western Inner Body, the Eastern Inner Body, the Foreground Component, and the Halo Candidate, close to the line-of-sight values except for the Foreground Component.
Along the bulk-motion axis, the dispersions of the two inner bodies and the Halo Candidate are larger (30.4, 29.8, and 59.3~\kms), so their plane-of-sky velocity ellipses are elongated, with axis ratios of 0.61--0.67 and major axes at position angles of 122$\degr$--136$\degr$, close to the bulk-motion axis.
The Foreground Component's ellipse, in contrast, is nearly round (axis ratio 0.86), with equal dispersions along and across the axis.
An excess along the bulk-motion axis can come from a spread in distance as well as in velocity.
If the intrinsic dispersions were the same in both directions, the excesses would correspond to depth dispersions of about 6.5, 3, and 2.5~kpc for the Halo Candidate, the Western Inner Body, and the Eastern Inner Body, comparable to the limits from the TRGB (Section~\ref{sec:distances}); the data therefore do not separate a spread in depth from an anisotropic velocity distribution.
In every direction, the Halo Candidate is the hottest component, even when the 9\% uncertainty of its distance is included.

\subsubsection{Motions relative to the Halo Candidate}

\begin{figure}
    \includegraphics[width=\columnwidth]{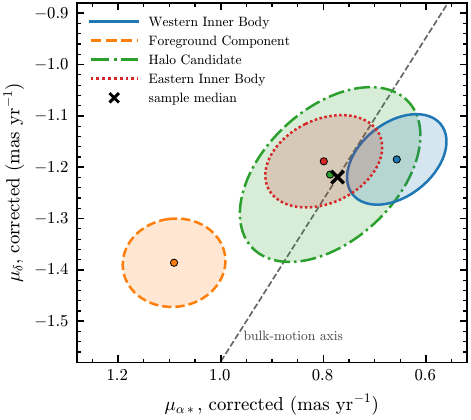}
    \caption{Mean perspective-corrected proper motions of the four components (points; all hard members) and the intrinsic 1$\sigma$ dispersions of their stars (ellipses, with measurement errors removed), with east to the left. The cross marks the median of the sample. The dashed line is the bulk-motion axis, the direction of the SMC's bulk proper motion, along which a difference in distance and a difference in motion produce the same proper motion. The motions relative to the Halo Candidate in \autoref{tab:kin} are measured against its stars in the same 1.5$\degr$ cells, so they differ from the differences between these means.}
    \label{fig:relmotion}
\end{figure}

\autoref{fig:relmotion} shows the mean proper motion of each component and the intrinsic spread of its stars.
The Foreground Component stands apart: its mean lies 0.36~mas~yr$^{-1}$ from the median of the sample, 0.31~mas~yr$^{-1}$ of it along the bulk-motion axis.
To separate motion from distance, we measure the motion of each component relative to the Halo Candidate, which we use as the reference because it is the broadest and most extended component and its mean motion is closest to the median of the sample (\autoref{fig:dist4}).
The relative motions across the bulk-motion axis and along the line of sight (\autoref{tab:kin}) cannot come from a difference in distance.
The bulk velocity field changes across the face of the SMC (Section~\ref{sec:exploratory}), so we compare each component with Halo Candidate stars in the same 1.5$\degr$ cells (22--32 cells per component), combining the cells weighted by their numbers of stars; changing the cell size or the position of the grid changes the results by at most 3~\kms.
The Foreground Component approaches us by $12.3\pm2.1$~\kms\ relative to Halo Candidate stars in the same cells and moves $30.4\pm2.6$~\kms\ across the bulk-motion axis, toward the northeast (position angle 58$\degr$).
Together, these give a lower limit of 33~\kms\ for its three-dimensional motion relative to the Halo Candidate.
Measured instead against all Halo Candidate stars, regardless of position, the same two offsets are larger, 22 and 38~\kms, because the Foreground Component lies on the eastern side, where the motion of the Halo Candidate differs from its average; along the line of sight, for example, velocities are lower on the eastern side (Section~\ref{sec:exploratory}).
The Western Inner Body moves $27.6\pm2.2$~\kms\ across the bulk-motion axis in the opposite direction (toward position angle 238$\degr$, i.e., southwest), with almost no offset along the line of sight ($-3.7\pm2.1$~\kms), while the Eastern Inner Body recedes by $19.9\pm2.1$~\kms\ and moves only $6.3\pm2.2$~\kms\ across the axis.
Because the two inner bodies overlap strongly, differences measured from their hard members are likely overestimates: in our mock catalogs with two overlapping populations, an offset of 29~\kms\ between the input populations is measured as 44~\kms\ from the hard members (Appendix~\ref{app:null}).

Along the bulk-motion axis, the relative motions depend on the distances, and ours do not constrain them: the 9\% uncertainty of the Halo Candidate's TRGB distance alone corresponds to about 40~\kms\ along this axis.
The Foreground Component and the Halo Candidate differ by $-18\pm41$~\kms\ along it, so the smaller distance of the Foreground Component can account for its large proper-motion offset along the axis (\autoref{fig:relmotion}) without any additional motion, although a motion of a few tens of \kms\ cannot be ruled out.
The differences for the two inner bodies ($-18\pm42$ and $-41\pm41$~\kms) are equally unconstrained.

\subsubsection{Line-of-sight velocity gradients}

All four components show a gradient in line-of-sight velocity across the sky (\autoref{tab:kin}; Section~\ref{sec:comp_maps}).
The gradients of the Western and Eastern Inner Bodies are the strongest, $5.1\pm0.2$ and $5.3\pm0.3$~\kms~deg$^{-1}$, with the velocity increasing toward position angles of 301$\degr$ and 296$\degr$, within 6$\degr$ of the direction away from the LMC (which lies at a position angle of 115$\degr$).
The Foreground Component has a gradient of $3.3\pm0.3$~\kms~deg$^{-1}$ toward 314$\degr$, and the Halo Candidate one of $3.4\pm0.5$~\kms~deg$^{-1}$ toward 271$\degr$, the most misaligned with the direction of the LMC (25$\degr$).
We discuss what these gradients may reflect in Section~\ref{sec:discussion}.

\subsubsection{Internal proper motions}

\begin{figure*}
    \includegraphics[width=\textwidth]{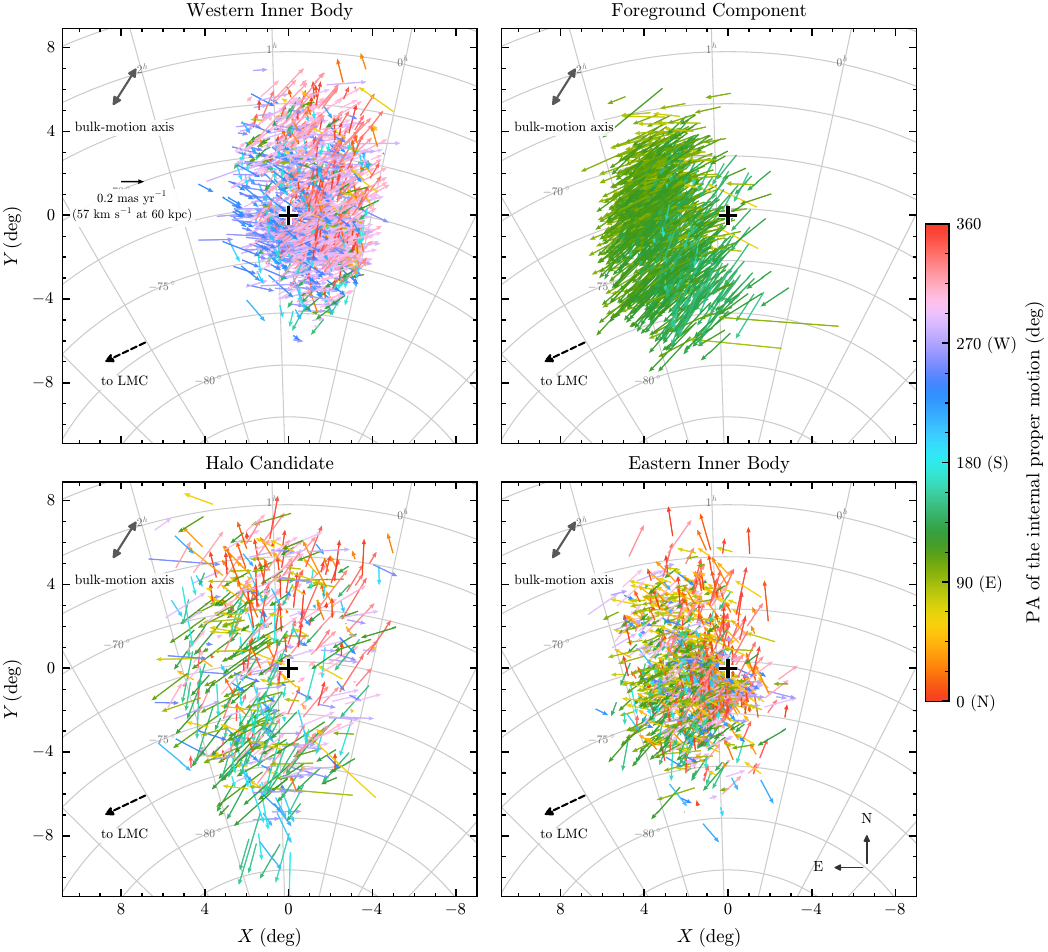}
    \caption{Internal proper motions of the four components: the perspective-corrected proper motion of each star minus the median of the sample, drawn in the projected $(X,Y)$ frame with east to the left (half of the stars of each component, chosen at random, are shown). Arrows are colored by the position angle of the motion, east of north. In each panel, the dashed arrow points toward the LMC and the solid double arrow marks the bulk-motion axis, along which a difference in distance and a difference in motion produce the same proper motion (Section~\ref{sec:kinematics}). The key gives the scale, with the equivalent velocity at 60~kpc. Thin gray lines mark right ascension (every 1$^{\rm h}$) and declination (every 2.5$\degr$).}
    \label{fig:quiver}
\end{figure*}

\autoref{fig:quiver} shows the internal proper motion of each star relative to the median of the sample.
The components show distinct, spatially coherent patterns.
The Foreground Component is the most coherent: nearly all of its stars move toward the east-southeast (mean position angle 118$\degr$), close to the direction of the LMC.
The Western Inner Body moves on average toward the west-northwest (287$\degr$), and the Halo Candidate and, more weakly, the Eastern Inner Body show an ordered north--south pattern.
Two effects limit how these patterns can be read.
First, the arrows are measured relative to the median of the sample, which lies between the components, so part of the opposition between the Foreground Component and the Western Inner Body follows from that choice.
Second, the component of each arrow along the bulk-motion axis (marked in each panel) mixes distance and motion.
Most of the Foreground Component's offset lies along this axis, where its smaller distance can account for it, as shown above; its apparent motion toward the LMC can therefore be largely a distance effect, while its motion across the axis, 30~\kms\ toward the northeast relative to Halo Candidate stars in the same cells (\autoref{tab:kin}), is the part that distance cannot produce.
The north--south pattern of the Halo Candidate lies roughly along the bulk-motion axis, so proper motions alone cannot distinguish a gradient in depth from an expansion.

\subsection{Chemistry}
\label{sec:chemistry}

The Halo Candidate is the most metal-poor component (Section~\ref{sec:overview}), and its metallicity distribution is also the broadest: its 16th--84th percentile range spans 0.52~dex, against 0.32--0.39~dex for the other three components.
It holds nearly all the metal-poor stars of our sample: 499 of the 524 stars with [Fe/H]$_{\rm corr}<-1.6$ are Halo Candidate members, while in each of the other components fewer than 2\% of the stars lie below $-1.5$.

\autoref{fig:radial}(b) shows how the median metallicity of each component changes with distance from its own centroid.
Both inner bodies become more metal-poor outward: the Eastern Inner Body from $-0.96$ at 0.6$\degr$ to $-1.15$ at 4.6$\degr$ ($-0.050\pm0.003$~dex~deg$^{-1}$ from a linear fit to its stars), and the Western Inner Body from $-1.03$ to $-1.15$ over a similar range ($-0.032\pm0.002$~dex~deg$^{-1}$).
The Foreground Component is nearly uniform in metallicity ($-0.011\pm0.004$~dex~deg$^{-1}$), and so is the Halo Candidate beyond 2$\degr$ from its centroid, where its median stays between $-1.58$ and $-1.54$ ($+0.006\pm0.007$~dex~deg$^{-1}$).
Closer to its centroid, the median of the Halo Candidate rises to $-1.43$.
This rise may reflect the overlap with the inner bodies in the hard assignment: within 1.5$\degr$ of the SMC center, its members with a membership probability below 0.9 have a median of $-1.18$, like the inner bodies, while those above 0.9 have $-1.75$.

\begin{figure*}
    \includegraphics[width=\textwidth]{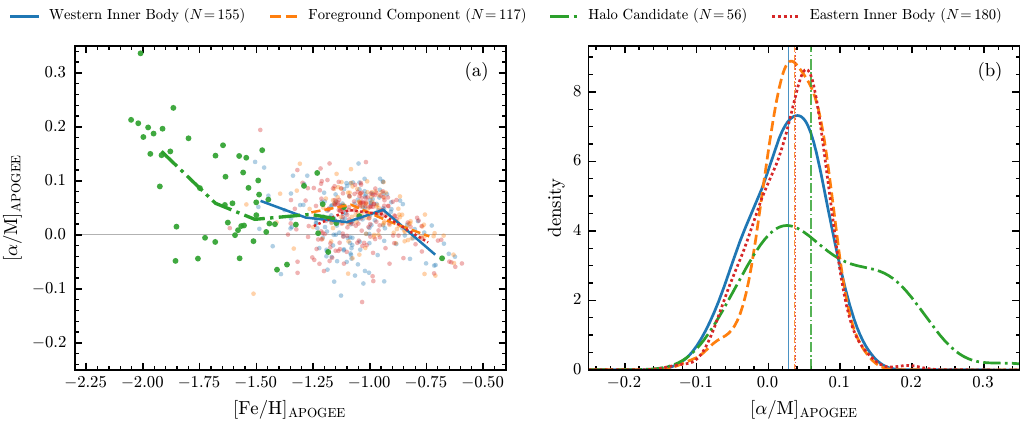}
    \caption{APOGEE DR17 abundances of the four components, for the 508 stars in common with our sample (S/N\,$>$\,20; hard members). (a) [$\alpha$/M] against [Fe/H]; Halo Candidate stars are drawn on top. Lines are running medians in 0.2~dex bins of [Fe/H] containing at least four stars. (b) Distributions of [$\alpha$/M] (kernel density estimates); vertical lines mark the medians. The legend gives the number of stars of each component.}
    \label{fig:alpha}
\end{figure*}

The APOGEE abundances, which do not enter the fit, test whether the components also differ in their chemical enrichment histories.
\autoref{fig:alpha} shows APOGEE [$\alpha$/M] against [Fe/H] for the 508 stars in common with our sample (S/N\,$>$\,20; Section~\ref{sec:feh_calib}).
For these stars, the median [Fe/H]$_{\rm corr}$ of each component agrees with its APOGEE median to within 0.01~dex.
All four components are consistent with one sequence, along which [$\alpha$/M] increases toward low metallicity.
In these stars, the median [$\alpha$/M] is nearly constant, $+0.03$ to $+0.04$, between [Fe/H]~$=-1.6$ and $-0.8$, and rises to $+0.15$--$0.21$ below $-1.8$ (14 stars). Larger APOGEE samples of the SMC show a similar rise toward lower metallicity, in [$\alpha$/Fe] \citep{nidever2020} and in [O/Fe], [Mg/Fe], [Si/Fe], and [Ca/Fe] \citep{hasselquist2021}.
The Halo Candidate has the highest median [$\alpha$/M] ($+0.059$, from 56 stars, against $+0.028$, $+0.037$, and $+0.038$ for the Western Inner Body, the Foreground Component, and the Eastern Inner Body), but its difference from the other three ($+0.026$, with a 95\% bootstrap range of $-0.010$ to $+0.065$) comes from its extension to lower metallicity.
Its 25 stars with [Fe/H]\,$<-1.6$, a range where the other components have one star, have a median of $+0.15$, while at fixed metallicity ($-1.5\leq$\,[Fe/H]\,$<-1.0$; 18 Halo Candidate stars and 287 others) the difference is $-0.009$ (95\% range $-0.028$ to $+0.030$).
The same holds for [$\alpha$/Fe]\,$=$\,[$\alpha$/M]\,$+$\,[M/H]\,$-$\,[Fe/H] and for O, Mg, Si, and Ca separately: at fixed metallicity, all the differences lie within 0.03~dex of zero, and none is significant.
We detect no offset at fixed metallicity: within the uncertainties, the Halo Candidate extends the sequence of the other components to lower metallicity rather than following a separate one.

\subsection{Maps of each component}
\label{sec:comp_maps}

\begin{figure*}
    \includegraphics[width=\textwidth]{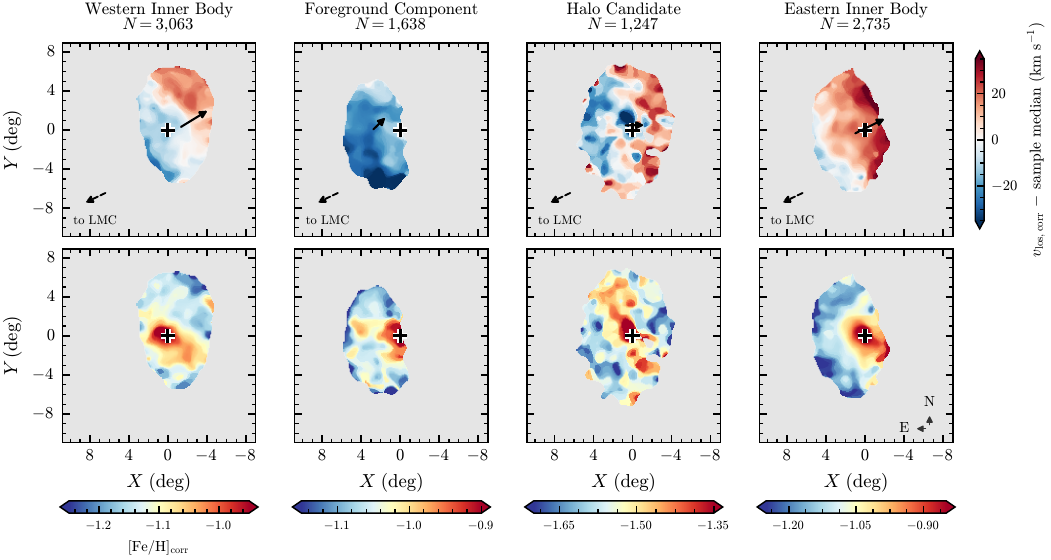}
    \caption{Line-of-sight velocity (top row; relative to the median of the whole sample) and metallicity (bottom row) of each component, smoothed as in \autoref{fig:maps_slices}: a Gaussian kernel of $\sigma=0.4\degr$ evaluated on 0.1$\degr$ pixels, showing the kernel-weighted median of the component's stars within $3\sigma$ of each pixel, where the kernel weights sum to at least five. The velocity maps share one color scale; each metallicity map has its own, spanning the 5th to 95th percentiles of its pixels. In the top row, the solid arrow starts at the component's centroid and points in the direction in which the velocity increases (\autoref{tab:kin}), with a length of 0.8$\degr$ per \kms~deg$^{-1}$; the dashed arrow points toward the LMC. Coordinates are the projected $(X,Y)$ frame, with east to the left; the cross marks the adopted SMC center.}
    \label{fig:comp_maps}
\end{figure*}

\autoref{fig:comp_maps} shows the line-of-sight velocity and the metallicity of each component across the face of the SMC, smoothed as in \autoref{fig:maps_slices}; the velocity maps share one color scale, and each metallicity map has its own.

In velocity, the gradients of \autoref{tab:kin} appear as smooth changes across each inner body: velocities rise from the southeast of the Western Inner Body to its northwest, and in the Eastern Inner Body the highest velocities lie along its western edge.
The velocities of the Foreground Component lie below the median of the sample everywhere (between $-38$ and $-8$~\kms\ over 90\% of its map) and are lowest in the south.
The map of the Halo Candidate, with its large dispersion and low density, is dominated by small-scale fluctuations.

In metallicity, both inner bodies have metal-rich centers, with [Fe/H]$_{\rm corr}\approx-0.9$ to $-1.0$, and more metal-poor outskirts, near $-1.2$.
In the Eastern Inner Body, the metal-rich region extends southwest of the center, along the SMC bar (Section~\ref{sec:exploratory}), so its stars west of its centroid are more metal-rich than those east of it (median $-0.98$ against $-1.11$).
The metallicity of the Foreground Component spans $-1.13$ to $-0.94$ over 90\% of its map, with its most metal-rich part at its western edge, closest to the SMC center; its color scale is the narrowest of the four, so these small variations stand out.
The Halo Candidate is metal-poor across its whole extent ($-1.67$ to $-1.40$ over 90\% of its map), except for a metal-rich patch at the center, which probably consists of inner-body stars erroneously assigned to it (Section~\ref{sec:chemistry}).
Beyond 2$\degr$ from its centroid, its median metallicity changes by less than 0.05~dex (\autoref{fig:radial}b).

\section{Summary and discussion}
\label{sec:discussion}

The large line-of-sight depth of the SMC has often been read as that of a single, highly inclined and disturbed galaxy \citep[e.g.,][and references therein]{kunkel2000}.
In our earlier APOGEE study \citep{almeida2024}, we found instead that on the eastern side much of this depth comes from a foreground population with the chemistry of the inner SMC.
Here we have decomposed 8,683 red giants across the SMC by their line-of-sight velocities, proper motions, and metallicities, ignoring their positions, and we find the following.
\begin{enumerate}
\item Four components describe the sample well, with three as a simpler alternative that merges the inner body (the bar and the surrounding main body) into one component. BIC prefers four to three (BIC$(K{=}3)-$BIC$(K{=}4)=155$), and none of the tested single-body models with smooth mean gradients, a constant intrinsic covariance, and Gaussian or moderately heavy tails reproduces this difference (at most 34 in 120 mock catalogs); a single body whose intrinsic spreads vary across it reaches it in 2 of 40 (Section~\ref{sec:choosing_k}). The inner body is therefore not one population with constant spreads; two of the four components are kinematical components of it, either two overlapping populations or one body in two kinematical states.
\item The Foreground Component (19\% of the sample) lies in front of the inner body, at $50.3\pm1.0$~kpc, and is thin along the line of sight (depth dispersion below 1.2~kpc). It approaches us relative to Halo Candidate stars at the same positions ($-12.3\pm2.1$~\kms) and moves $30.4\pm2.6$~\kms\ across the bulk-motion axis, at least 33~\kms\ in three dimensions. It is the coldest component along the line of sight, has the metallicity of the inner bodies, and has a red giant branch similar to theirs.
\item The Halo Candidate is the most metal-poor component (median [Fe/H]$_{\rm corr}=-1.54$), the hottest in every direction ($\sigma_{\rm los}=37.9$~\kms), and the most extended; its mean proper motion is close to that of the sample, and its [$\alpha$/M] extends, within the uncertainties, the sequence of the other components to lower metallicity. Its red giant branch makes it the oldest component, about 1.8 times older than the inner bodies.
\item The two inner-body components have the same metallicity within 0.04~dex, TRGB distances that differ by only $3.7\pm2.1$~kpc, and indistinguishable red giant branches, but different mean motions.
\item Every component shows a line-of-sight velocity gradient. Those of the two inner-body components are the strongest, $\approx$5~\kms~deg$^{-1}$, and point within 6$\degr$ of the direction away from the LMC.
\end{enumerate}

\subsection{The Foreground Component: material pulled out of the SMC}
\label{sec:disc_fg}

The Foreground Component corresponds to the near population on the eastern side of the SMC that \citet{almeida2024} identified using APOGEE data. It also corresponds to the near component of the distance bimodality that red clump stars show there \citep{nidever2013,subramanian2017,tatton2021,omkumar2021}. Red giants on that side have a bimodal line-of-sight velocity distribution \citep{james2021}, whereas RR Lyrae stars apparently show no clear distance bimodality \citep{muraveva2018}.
Here the Foreground Component emerges as a distinct entity, with 1,638 stars and a distance from its own TRGB.

Its properties point to stars pulled out of the SMC during its recent interaction with the LMC.
The Foreground Component lies in front of the inner body on its LMC-facing side, approaches us relative to Halo Candidate stars at the same positions, and moves across the bulk-motion axis, where a smaller distance adds no contribution from the SMC's bulk motion.
Its metallicity and red giant branch resemble those of the inner bodies, and its [$\alpha$/M] follows the same sequence (\autoref{fig:alpha}).

The case for an SMC rather than an LMC origin rests on the kinematics and the color--magnitude diagram, not on the metallicity: stellar debris from the LMC near the SMC spans a similar range of metallicity \citep{massana2024}.
The line-of-sight velocity of the Foreground Component lies within about 12~\kms\ of that of the Halo Candidate at the same positions. For comparison, the LMC's systemic line-of-sight velocity is more than 100~\kms\ higher than the SMC's \citep{vandermarel2002,harris2006}. \citet{massana2024} find the debris immediately around the SMC to be mostly SMC material, with some from the LMC.
Stars from the inner SMC are not what gradual tidal stripping, which removes the least bound outer stars first, would produce.
As argued by \citet{almeida2024}, they point instead to a close encounter with a small impact parameter \citep{besla2012,zivick2018}, which can disturb the inner SMC directly \citep{rathore2025}; \citet{omkumar2021} likewise suggested that their red-clump foreground might have originated in the inner SMC.

The geometry of the Foreground Component supports this reading.
The edge of its luminosity function is the sharpest of the four components, so its depth dispersion is below 1.2~kpc (Section~\ref{sec:distances}), while on the sky it extends over $2.5\degr\times1.3\degr$ (1$\sigma$; $\approx$2.2$\times$1.1~kpc at its distance), elongated north--south, about 75$\degr$ from the direction of the LMC.
In an exploratory split into northern and southern halves, and into eastern and western halves, the TRGB distances of the halves differ by only $-0.6\pm1.1$ and $-1.0\pm0.9$~kpc, so the component is a thin sheet roughly perpendicular to the line of sight.
In the same split, the southern half moves faster than the northern half along the bulk-motion axis, by more than the distance differences allowed by these uncertainties can produce; the implied stretching rate, 7--9~\kms~kpc$^{-1}$, corresponds to $\approx$110--135~Myr.
This rests on a single split of hard members, and correlated \gaia\ proper-motion errors reach $\approx$0.02~mas~yr$^{-1}$ on degree scales, so we regard this calculation as indicative only.
Whether the sheet is still bound to the SMC is not determined by our data.
It lies farther in front of the inner body than the SMC's tidal radius estimated by \citet[][3.4--5.1~kpc]{almeida2024}, but deciding whether its stars are bound requires dynamical models.

\subsection{The Halo Candidate: an extended, metal-poor SMC component}
\label{sec:disc_halo}

The Halo Candidate has the properties expected of a stellar halo of the SMC: it is the most metal-poor, hottest, most extended, and oldest component, and it is centered like the sample as a whole.
\citet{almeida2024} found more metal-poor stars in the western and far-eastern periphery of the SMC, but read them as the periphery of the main body; with a sample about eight times larger, those stars now separate as a component of their own.
RR Lyrae stars, which trace the oldest population, form an ellipsoidal distribution with a line-of-sight depth dispersion of 5.2~kpc for the whole sample and 1--10~kpc in individual regions \citep{muraveva2018}; the TRGB edge of the Halo Candidate allows a depth dispersion of this size (below 9.9~kpc, 1$\sigma$).

We call this grouping of stars a ``halo candidate'' for three reasons.
First, its hard members are a selection: where it overlaps the inner bodies, its stars with velocities close to theirs are preferentially assigned to them, which pushes its median metallicity and dispersion toward the extreme. Near the center, the same overlap can leave inner-body-like stars among its members, which probably produces the metal-rich patch there (Section~\ref{sec:chemistry}).
Second, its chemistry does not establish its origin: it follows the sequence of the other components, as expected of SMC material, but the 56 stars with APOGEE abundances cannot exclude debris from an early accretion of a chemically similar system.
Third, Milky Way giants that pass the survey selection would preferentially be assigned to it; if all the expected ones were assigned to it, they would make up about 8\%--9\% of its members (Section~\ref{sec:overview}).
Nor is it unaffected by the interaction: it shows a line-of-sight velocity gradient of 3.4~\kms~deg$^{-1}$, although one less aligned with the direction of the LMC (25$\degr$) than those of the inner bodies.

\subsection{The inner body: one body or two populations?}
\label{sec:disc_inner}

In the adopted four-component model, the inner body is described by two components, but the model selection does not determine whether they are two overlapping populations or one body whose kinematical state changes across it (Section~\ref{sec:choosing_k}).
The two components have the same metallicity within 0.04~dex, TRGB distances that differ by only $3.7\pm2.1$~kpc, and indistinguishable red giant branches (4.8 and 4.7~Gyr).
They differ in position and motion: the Western Inner Body is centered 1.0$\degr$ west of the center and moves toward the southwest across the bulk-motion axis, while the Eastern Inner Body is centered 1.4$\degr$ east of it and recedes from us, both relative to the Halo Candidate.
We interpret them as one inner body that responds differently across its extent to the interaction with the LMC.
Because two populations of one galaxy could also share their chemistry, distance, and age, this remains an interpretation.
Other tracers also show structure consistent with tidal disturbance of the SMC: classical Cepheids have a three-dimensional distribution elongated over more than 25~kpc from the east and northeast to the southwest, with the older Cepheids more extended than the younger ones \citep{ripepi2017}, and intermediate-age stars show irregular central features suggesting that tidal interactions have also affected the inner SMC \citep{elyoussoufi2019}.

The line-of-sight velocity gradients of both inner-body components point within 6$\degr$ of the direction away from the LMC, so that velocities increase away from it.
If the LMC-facing side is nearer, as the TRGB distances suggest, this is the sense expected for stretching along the LMC--SMC axis, which the SMC's proper motions also show \citep{zivick2018,zivick2021,vijayasree2026}.
We detect no rotation signature.
We fit a linear field to the three velocity components of each inner body, placing all the stars at a common distance, an approximation that keeps errors in individual distances from mimicking rotation.
The proper-motion field then diverges rather than circulates, and a rotating, inclined solid-body disk fitted to these fields is strongly disfavored.
Any rotation of the old stars in the plane of the sky is below $\approx$5--8~\kms.
This agrees with the absence of rotation in the VMC proper motions \citep{vijayasree2026} and with collision models in which the old stars are dispersion-dominated and retain at most $\approx$10~\kms\ of rotation within $\approx$1~kpc of the center \citep{rathore2025}.

\subsection{The line-of-sight structure of the SMC}
\label{sec:disc_picture}

Together, the components describe the SMC as several stellar populations in different kinematical states that overlap on the sky and along the line of sight, rather than as a single inclined, rotating disk.
A thin foreground sheet of inner-SMC stars lies on the LMC-facing side, an inner body shows velocity gradients along the LMC axis and no detectable rotation, and an extended, metal-poor, dynamically hot component surrounds them.
Part of the SMC's line-of-sight depth is therefore the separation between components rather than the thickness of one body: the TRGB edges limit the depth dispersions of the two inner-body components to 3.4 and 5.1~kpc (1$\sigma$), while the Foreground Component lies 6.3--10.0~kpc in front of them.

The components also differ from the structures by which the SMC is usually described. The bar and the Wing stand out in young stars, which follow the irregular distribution of the H\,{\sc i} gas, whereas older stars trace a more regular, elliptical body \citep{elyoussoufi2019,elyoussoufi2023}, with two central overdensities along its northeast--southwest axis that may be a projection effect \citep{elyoussoufi2023}. Our components divide this older population by velocity, proper motion, and metallicity rather than by position: the two inner-body components share the central body, one on each side, while the Foreground Component and the Halo Candidate lie mostly outside it. Sky positions do not enter the fit, so this arrangement is not assumed by the model (Section~\ref{sec:gmm_idea}).

The star-forming gas of the SMC also shows two superimposed systems, about 5~kpc apart along the line of sight \citep{murray2024}.
Those authors discuss whether the more distant system is tidal debris behind the main body, the ``counter-bridge'' predicted by the simulations of \citet{diaz2012}.
Among the stars, the counter-bridge has been associated with Cepheids \citep{ripepi2017}, with distant red clump stars in the north \citep{tatton2021}, and with a northward motion \citep{niederhofer2021} that \citet{vijayasree2026} report only among the red giant branch stars in their sample and attribute to an interaction more than 2~Gyr ago.
Our Foreground Component is a different structure: it lies on the near side of the SMC, not behind it.

Models place the last LMC--SMC encounter within the past few hundred million years and at a small separation: 147~$\pm$~33~Myr ago at $7.5\pm2.5$~kpc from the SMC's proper motions \citep{zivick2018}, or a direct collision about 100~Myr ago at an impact parameter of about 2~kpc \citep{rathore2025}.
The characteristic stretching timescale of the Foreground Component, $\approx$110--135~Myr, is of this order, as expected if it was pulled out in that encounter.
A collision about 140~Myr ago has also been proposed, still preliminarily, to explain the warp of the outer LMC disk \citep{oden2026b}.

These results have two consequences for models of the SMC.
First, a foreground sheet that holds about a fifth of the red giants of our sample must be part of any description of the SMC's structure: fits of a single inclined disk to distance indicators on the eastern side partly measure the separation between components.
Second, the velocity gradients along the LMC axis, the absence of detectable rotation, and the components in different kinematical states indicate substantial tidal disturbance and make equilibrium assumptions uncertain, so masses from rotation curves or virial estimators should be treated with caution \citep{rathore2025}.

\subsection{Scope of the decomposition}
\label{sec:caveats}

Our components describe the joint distribution of four quantities.
Their memberships are probabilistic, so where two components overlap, hard assignment separates their properties further than the underlying populations are separated (Sections~\ref{sec:fitting} and \ref{sec:kinematics}).
The sample also inherits the selection of the survey (Section~\ref{sec:mgs}) and its footprint, which ends at RA~$=0^{\rm h}$.
The photometry behind the TRGB distances and the APOGEE abundances do not enter the fit, which makes them useful checks on the components, although comparing them between components uses the fitted memberships.

\section{Conclusions and future work}
\label{sec:conclusions}

With a Gaussian mixture of line-of-sight velocity, proper motion, and metallicity that accounts for each star's uncertainties, we have separated 8,683 SMC red giants into four components: a thin foreground sheet of inner-SMC stars on the LMC-facing side, with a distinct motion relative to the Halo Candidate; two kinematical components of the inner body, with similar metallicities and inferred ages and distances consistent within the uncertainties, but different motions; and an extended, metal-poor, dynamically hot halo candidate.
Part of the SMC's line-of-sight structure is the superposition of these components, which models of the LMC--SMC interaction need to reproduce.

These results suggest several next steps.
\begin{itemize}
\item[$\bullet$] \textbf{Ages.} Main-sequence turnoffs from deep photometry, from SMASH \citep{nidever2021}, potentially from Euclid imaging of the SMC \citep{scaramella2022}, and later from LSST \citep{ivezic2019} and the Nancy Grace Roman Space Telescope \citep[formerly WFIRST;][]{akeson2019}, will better constrain the ages of the Halo Candidate and help resolve the age--metallicity degeneracy of fits to the red giant branch.
\item[$\bullet$] \textbf{Distances.} Red clump and RR Lyrae stars already map the SMC in three dimensions \citep{muraveva2018,tatton2021}, but these distance tracers have not been linked to the kinematical components identified here. Spectra of these stars, planned by 4MOST \citep{cioni2019}, would help associate them with the components and constrain each one's three-dimensional rotation and expansion.
\item[$\bullet$] \textbf{Models.} Comparing the components with simulations of the LMC--SMC interaction, observed with the selection of our survey, will test whether the Foreground Component is bound and when it was pulled out.
\item[$\bullet$] \textbf{Chemistry.} More precise abundances for more stars will test whether the stars of the Halo Candidate formed within the SMC or were accreted from another system.
\item[$\bullet$] \textbf{Sample.} Larger samples from SDSS-V and from the One Thousand and One Magellanic Fields survey of 4MOST, which will obtain spectra of about half a million stars of many populations across both Clouds \citep{cioni2019}, will refine these components and test them with more stars. \gaia\ DR4, due in December 2026, will extend the proper-motion baseline and provide low-resolution spectra for more stars, from which metallicities can be estimated.
\end{itemize}
The method, a mixture with measurement uncertainties fitted without positions and validated with mock catalogs, can also be applied to the other dwarf galaxies that upcoming spectroscopic surveys will map.

\section*{Data Availability}
The \gaia~DR3 astrometry and photometry used in this work are publicly available from the \gaia~Archive (\url{https://gea.esac.esa.int/archive/}), and the APOGEE DR17 abundances used for the metallicity recalibration are publicly available through the SDSS-IV DR17 data archive (\url{https://www.sdss4.org/dr17/}). The SDSS-V/BOSS spectra of the Magellanic Genesis Survey and the MINESweeper stellar parameters derived from them are not yet public and will become public in a forthcoming SDSS-V data release. The value-added catalog underlying this analysis---the 8,683 RGB stars with their recalibrated metallicities, corrected line-of-sight velocities and proper motions, distance proxies, and $K=4$ membership probabilities and hard assignments---will be made public in machine-readable form once those data are released.

\section*{Acknowledgments}
We thank Knut A. G. Olsen for providing the center-of-mass and perspective corrections used in Section~\ref{sec:com_frame}.
A.A. and S.R.M. acknowledge support from NSF grants AST 1909497 and AST 2408160. S.H. was supported by the DDRF grant D0001.82510. D.L.N. acknowledges support from NSF grants AST 1908331 and 2408159. D.H. is supported by the UKRI Science and Technology Facilities Council under project 101148371 as a Marie Sk\l{}odowska-Curie Research Fellowship. B.D. acknowledges support from ANID Basal project FB210003 and also that this work was funded by the 2026 Regular Bridge Fund of the Universidad Andr\'es Bello (UNAB), project No.~DI-01-26/FPFR. This work was partially funded by the National Science Foundation's support of the Sloan Digital Sky Survey's FAST 4ward program through award AST-2425222. Support for this work was provided by NASA through the NASA Hubble Fellowship grant HST-HF2-51600.001-A awarded by the Space Telescope Science Institute, which is operated by the Association of Universities for Research in Astronomy, Incorporated, under NASA contract NAS5-26555.

Funding for the Sloan Digital Sky Survey V has been provided by the Alfred P. Sloan Foundation, the Heising-Simons Foundation, the National Science Foundation, and the Participating Institutions. SDSS acknowledges support and resources from the Center for High-Performance Computing at the University of Utah. SDSS telescopes are located at Apache Point Observatory, funded by the Astrophysical Research Consortium and operated by New Mexico State University, and at Las Campanas Observatory, operated by the Carnegie Institution for Science. The SDSS web site is \url{www.sdss.org}.

SDSS is managed by the Astrophysical Research Consortium for the Participating Institutions of the SDSS Collaboration, including the Carnegie Institution for Science, Chilean National Time Allocation Committee (CNTAC) ratified researchers, Caltech, the Gotham Participation Group, Harvard University, Heidelberg University, The Flatiron Institute, The Johns Hopkins University, L'Ecole polytechnique f\'{e}d\'{e}rale de Lausanne (EPFL), Leibniz-Institut f\"{u}r Astrophysik Potsdam (AIP), Max-Planck-Institut f\"{u}r Astronomie (MPIA Heidelberg), Max-Planck-Institut f\"{u}r Extraterrestrische Physik (MPE), Nanjing University, National Astronomical Observatories of China (NAOC), New Mexico State University, The Ohio State University, Pennsylvania State University, Smithsonian Astrophysical Observatory, Space Telescope Science Institute (STScI), the Stellar Astrophysics Participation Group, Universidad Nacional Aut\'{o}noma de M\'{e}xico, University of Arizona, University of Colorado Boulder, University of Illinois at Urbana-Champaign, University of Toronto, University of Utah, University of Virginia, Yale University, and Yunnan University.

Funding for the Sloan Digital Sky Survey IV has been provided by the Alfred P. Sloan Foundation, the U.S. Department of Energy Office of Science, and the Participating Institutions.

SDSS-IV acknowledges support and resources from the Center for High Performance Computing at the University of Utah. The SDSS website is www.sdss4.org.

SDSS-IV is managed by the Astrophysical Research Consortium for the Participating Institutions of the SDSS Collaboration including the Brazilian Participation Group, the Carnegie Institution for Science, Carnegie Mellon University, Center for Astrophysics $|$ Harvard \& Smithsonian, the Chilean Participation Group, the French Participation Group, Instituto de Astrof\'isica de Canarias, The Johns Hopkins University, Kavli Institute for the Physics and Mathematics of the Universe (IPMU) / University of Tokyo, the Korean Participation Group, Lawrence Berkeley National Laboratory, Leibniz Institut f\"ur Astrophysik Potsdam (AIP), Max-Planck-Institut f\"ur Astronomie (MPIA Heidelberg), Max-Planck-Institut f\"ur Astrophysik (MPA Garching), Max-Planck-Institut f\"ur Extraterrestrische Physik (MPE), National Astronomical Observatories of China, New Mexico State University, New York University, University of Notre Dame, Observat\'ario Nacional / MCTI, The Ohio State University, Pennsylvania State University, Shanghai Astronomical Observatory, United Kingdom Participation Group, Universidad Nacional Aut\'onoma de M\'exico, University of Arizona, University of Colorado Boulder, University of Oxford, University of Portsmouth, University of Utah, University of Virginia, University of Washington, University of Wisconsin, Vanderbilt University, and Yale University.

This work has made use of data from the European Space Agency (ESA) mission
{\it Gaia} (\url{https://www.cosmos.esa.int/gaia}), processed by the {\it Gaia}
Data Processing and Analysis Consortium (DPAC,
\url{https://www.cosmos.esa.int/web/gaia/dpac/consortium}). Funding for the DPAC
has been provided by national institutions, in particular the institutions
participating in the {\it Gaia} Multilateral Agreement.

This publication makes use of data products from the Two Micron All
Sky Survey \citep{skrutskie2006}, which is a joint project of the University of Massachusetts
and the Infrared Processing and Analysis Center/California Institute of
Technology, funded by the National Aeronautics and Space Administration
and the National Science Foundation.

This publication makes use of data products from the Wide-field Infrared Survey Explorer \citep{wright2010}, which is a joint project of the University of California, Los Angeles, and the Jet Propulsion Laboratory/California Institute of Technology, and NEOWISE, which is a project of the Jet Propulsion Laboratory/California Institute of Technology. WISE and NEOWISE are funded by the National Aeronautics and Space Administration. The WISE photometry used here is from the unWISE coadds and catalog \citep{meisner2019,schlafly2019}.

This research has made use of NASA’s Astrophysics Data System Bibliographic Services; the arXiv preprint server operated by Cornell University; and the SIMBAD databases hosted by the Strasbourg Astronomical Data Center.

\par\vspace{0pt plus 6\baselineskip}\penalty-250\vspace{0pt plus -6\baselineskip}
\section*{AI Disclosure}
AI tools were used in preparing the manuscript. Claude (Anthropic) was used to edit and format text, to check numbers in the text and tables against the analysis output, to optimize code for the figures, and to correct spelling. ChatGPT (OpenAI) was used to double-check calculations and to improve the clarity and grammar of earlier drafts. The authors reviewed all AI-assisted text, code, and checks, and take full responsibility for the content of the manuscript.

\bibliographystyle{aasjournal}
\bibliography{ref_v6.bib}

\appendix

\section{Alternative mixture size: $K=3$}
\label{app:k3}
\restartappendixnumbering

\begin{figure}[ht!]
    \centering
    \includegraphics[width=0.5\textwidth]{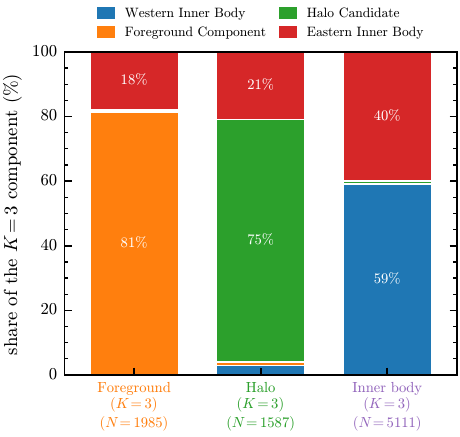}
    \caption{Composition of the three components of the $K=3$ model in terms of the components of the fiducial four-component model (hard members). Each bar is one $K=3$ component, labeled with its number of members; the colored segments give the share of its members that belong to each fiducial component.}
    \label{fig:k3}
\end{figure}

We fitted the same 8,683 stars with three components, using the same quantities, uncertainties, and fitting procedure as for the fiducial model (Section~\ref{sec:fitting}).
The fit converged after 31 iterations; repeating it from five starting points improves the log-likelihood by at most 0.12 and gives memberships with an adjusted Rand index of 0.965 relative to it.
We compare its hard members with those of the four-component model star by star (\autoref{tab:k3} and \autoref{fig:k3}).

The three-component model merges the two inner bodies: 99.0\% of the members of its inner-body component belong to the Western Inner Body (59.0\%) or the Eastern Inner Body (40.0\%).
Its other two components correspond to the Foreground Component and the Halo Candidate, which provide 81.3\% and 75.0\% of their members; most of the remaining members come from the Eastern Inner Body (18.0\% and 20.9\%).
The median metallicities of the corresponding components differ by at most 0.07~dex.

The Eastern Inner Body is the component whose stars are redistributed.
Of its 2,735 members, 74.8\% enter the merged inner-body component, 13.1\% join the Foreground component of the three-component model, and 12.1\% join its Halo component.
The stars that join the Halo component are the metal-poor part of the Eastern Inner Body (median [Fe/H]$_{\rm corr}=-1.27$, against $-1.01$ for those that stay in the merged component): without a fourth component, they are absorbed by the metal-poor component.

The two models therefore differ mainly in whether the inner body of the SMC is described by one component or two; the Foreground Component and the Halo Candidate are recovered with most of their members and with similar metallicities.

\startlongtable
\begin{deluxetable}{lccccc}
\tablecaption{Corresponding components of the three- and four-component models \label{tab:k3}}
\tablehead{
\colhead{Component} & \colhead{$N_{K=4}$} & \colhead{$N_{K=3}$} & \colhead{From counterpart (\%)} & \colhead{[Fe/H]$_{{\rm corr},K=4}$} & \colhead{[Fe/H]$_{{\rm corr},K=3}$}
}
\startdata
Foreground Component & 1,638 & 1,985 & 81.3 & $-1.04$ & $-1.06$ \\
Halo Candidate & 1,247 & 1,587 & 75.0 & $-1.54$ & $-1.47$ \\
Inner body\tablenotemark{a} & 5,798 & 5,111 & 99.0 & $-1.09$, $-1.05$ & $-1.06$ \\
\enddata
\tablenotetext{a}{The Western and Eastern Inner Bodies of the four-component model, merged into one component in the three-component model; the two values of [Fe/H]$_{{\rm corr},K=4}$ are theirs.}
\tablecomments{Hard members. From counterpart: the percentage of the members of the $K=3$ component that belong to its counterpart (or counterparts) in the four-component model. [Fe/H]$_{\rm corr}$: medians (Equation~\ref{eq:feh_calib}).}
\end{deluxetable}

\section{Choosing the number of components: details}
\label{app:modelsel}
\restartappendixnumbering

This appendix gives the details of the tests summarized in Section~\ref{sec:choosing_k}.
All fits use the configuration of Section~\ref{sec:fitting}; the number of starting points differs between the tests and is given for each.

\subsection{Fits and numerical checks}

The fiducial $K=4$ fit converged after 54 iterations.
Repeating it from 15 starting points improves the log-likelihood by at most 0.24 and gives memberships with an adjusted Rand index of 0.991 relative to the fiducial ones.
For each $K$ from 1 to 8 we kept the best of 15 fits.
The log-likelihoods of the 15 fits differ by at most 1.1 up to $K=6$, so the solutions are consistent across starting points there; the spread grows to 3.6 and 31 for $K=7$ and 8.
None of the fits used for the model-selection tests (the sweep in $K$, the bootstrap resamples, the gap-statistic references, and the synthetic test) reached the limit of 1,000 iterations, and none failed; the fits of the null test are described in Appendix~\ref{app:null}.

\subsection{Information criteria}

\autoref{tab:ic} lists the maximum log-likelihood, BIC, and gap statistic for $K=1$--8, with and without the measurement uncertainties.
It also lists the Akaike information criterion, $\mathrm{AIC}=2p-2\ln L_{\rm max}$ \citep{akaike1974}, which penalizes additional parameters less than BIC does for a sample of our size; AIC decreases steadily up to $K=8$, the largest value we tested, so we do not use it.
Without the measurement uncertainties, BIC is lowest at $K=4$ (higher by 41 for $K=3$ and by 33 for $K=5$); its minimum moves to $K=6$ only when the uncertainties are deconvolved.

\begin{deluxetable}{rrrrrcrc}
\tablecaption{Information criteria and gap statistic \label{tab:ic}}
\tablehead{
\colhead{$K$} & \colhead{$p$} & \colhead{$\ln L_{\rm max}$} & \colhead{$\Delta$BIC} & \colhead{$\Delta$AIC} & \colhead{Gap$(K)$} & \colhead{$\Delta$BIC, no errors} & \colhead{Gap$(K)$, no errors}
}
\startdata
1 & 14 & $-48366.7$ & 2169.1 & 2733.2 & $0.5151\pm0.0010$ & 2631.5 & $0.4998\pm0.0010$ \\
2 & 29 & $-47753.0$ & 1077.9 & 1535.9 & $0.5194\pm0.0009$ & 849.4 & $0.5109\pm0.0009$ \\
3 & 44 & $-47268.2$ & 244.3 & 596.3 & $0.5232\pm0.0009$ & 40.9 & $0.5150\pm0.0009$ \\
4 & 59 & $-47122.6$ & 89.1 & 335.1 & $0.5232\pm0.0011$ & 0.0 & $0.5131\pm0.0011$ \\
5 & 74 & $-47050.6$ & 81.0 & 220.9 & $0.5220\pm0.0009$ & 33.1 & $0.5117\pm0.0008$ \\
6 & 89 & $-46942.0$ & 0.0 & 33.9 & $0.5196\pm0.0021$ & 164.9 & $0.5094\pm0.0007$ \\
7 & 104 & $-46916.5$ & 85.0 & 12.8 & $0.5170\pm0.0025$ & 211.1 & $0.5079\pm0.0011$ \\
8 & 119 & $-46895.1$ & 178.2 & 0.0 & $0.5163\pm0.0032$ & 243.7 & $0.5076\pm0.0012$ \\
\enddata
\tablecomments{Fits with measurement uncertainties (\texttt{pyGMMis}; best of 15 starts per $K$) and without them (scikit-learn). $p=15K-1$ is the number of free parameters; $\ln L_{\rm max}$ is for the standardized quantities. $\Delta$BIC and $\Delta$AIC are relative to the minimum of each column. Gap$(K)$ is computed from single-start fits (Appendix~\ref{app:gap}).}
\end{deluxetable}

\subsection{Gap statistic}
\label{app:gap}

We use a likelihood-based adaptation of the gap statistic of \citet{tibshirani2001}, in which the deviance $W_K=-2\ln L$ of the $K$-component fit replaces their within-cluster dispersion:
\begin{equation}
\mathrm{Gap}(K)=\frac{1}{B}\sum_{b=1}^{B}\ln W^{*}_{K,b}-\ln W_K ,
\label{eq:gap}
\end{equation}
where $W^{*}_{K,b}$ is the deviance of the same model fitted to the $b$th of $B=100$ reference samples.
Because the deviance depends on the units of the data, we compute it for the standardized quantities of Section~\ref{sec:features}.
Each $W_K$ and $W^{*}_{K,b}$ comes from a single fit from the fiducial starting point, so $W_K$ can differ slightly from $-2\ln L_{\rm max}$ in \autoref{tab:ic}.
The reference samples contain no structure: their stars are drawn uniformly within the box that encloses the data along its principal axes \citep[the second method of][]{tibshirani2001}, and every reference star is given the median of the observed uncertainties.
A larger gap means that the $K$-component model captures more structure in the data than it finds in structureless samples.
The uncertainty of Gap$(K)$ is $s_K=\sigma_K\sqrt{1+1/B}$, where $\sigma_K$ is the standard deviation of $\ln W^{*}_{K,b}$ over the reference samples.
The usual selection rule, which we use only as a guide for this adapted statistic, takes the smallest $K$ whose gap is not significantly lower than that of the next one, $\mathrm{Gap}(K)\geq\mathrm{Gap}(K+1)-s_{K+1}$.
With the measurement uncertainties, this rule selects $K=3$, but Gap(3) and Gap(4) differ by only 0.03 times their uncertainty, so the two are indistinguishable, whereas Gap(2) is lower than Gap(3) by about four times the uncertainty of Gap(3).
Without the uncertainties the rule also selects $K=3$, with Gap(4) lower by 1.7 times its uncertainty.

\subsection{Reproducibility}

We drew 100 bootstrap resamples of the catalog and fitted each with $K=3$ to 7 (best of three starts per $K$).
For $K=3$ to 6 we compared the memberships that the fits to each pair of resamples assign to the stars of the catalog (4,950 pairs), using the adjusted Rand index.
Because the pairs share fits, they are not independent; we therefore estimate the uncertainties below by drawing the 100 bootstrap fits with replacement 2,000 times and recomputing the statistics.
The mean index is 0.865, 0.761, 0.598, and 0.605 for $K=3$, 4, 5, and 6 (standard errors of the mean 0.004, 0.008, 0.016, and 0.010); including the measurement uncertainties raises it by $\approx$0.03 at every $K$.
Compared on the same pairs of resamples, $K=4$ exceeds $K=6$ by $0.156\pm0.011$ (in 94\% of the pairs) and $K=5$ by $0.163\pm0.016$ (87\%), while $K=3$ exceeds $K=4$ by $0.104\pm0.009$; reproducibility alone would therefore favor $K=3$, whereas BIC favors $K=4$ over $K=3$.
Beyond $K=4$ the agreement falls to a plateau at $\approx$0.60, with no second, more reproducible solution at $K=5$ or 6.
Applied to each resample, BIC selects $K=4$, 5, 6, and 7 in 8, 28, 57, and 7 cases: its minimum moves among $K=4$ to 7, but the components it adds do not recur from one resample to the next.

\subsection{Synthetic catalog}

To test the method itself, we drew a synthetic catalog from the fitted $K=4$ model, with one synthetic star per real star scattered by that star's uncertainties, and fitted it exactly as the data.
The fit recovers the component means to within 1.3~\kms, 0.02~dex, and 0.016~mas~yr$^{-1}$, the intrinsic dispersions to within 1.3~\kms, 0.013~dex, and 0.011~mas~yr$^{-1}$, and the weights to within $\approx$0.02.
BIC selects the true $K=4$: BIC$(K{=}3)-$BIC$(K{=}4)=165$, and $K=5$ and 6 are higher than $K=4$ by 125 and 258.
Twenty further synthetic catalogs give BIC$(K{=}3)-$BIC$(K{=}4)$ between 115 and 216 (median 162), bracketing the value of 155 for the data.
The hard assignments are correct for 74\% of the synthetic stars (72\% if the uncertainties are ignored): with the overlaps of our components, about one star in four is assigned to the wrong component even when the model is exact.

\subsection{Null test for the inner body}
\label{app:null}

The null test asks whether a single inner body, fitted by the same procedure, would also be split into two components.
We replaced the 5,798 stars assigned to the western and eastern inner-body components with a single mock population, and kept the stars of the other two components as observed.
For each of the four quantities, we fitted a smooth mean field across the sky to the real inner-body stars (a quadratic function of $X$ and $Y$, with 3$\sigma$ clipping; a cubic function and averages in 1.5$\degr$ cells for comparison), and computed the intrinsic $4\times4$ covariance of the residuals about it by subtracting the mean measurement covariance.
Each mock star keeps the position and uncertainties of a real star. Its four values are the sum of the smooth mean field at that position, a random draw from the intrinsic distribution, and noise drawn from that star's uncertainties.
Variants use heavier-tailed intrinsic distributions (Student-$t$), or let the intrinsic covariance vary across the inner body (computed separately in six east--west strips or on a $3\times3$ grid).
Two-population controls draw the western and eastern components separately.
In the first (P2), each star is assigned to one of the two by drawing from its membership probabilities, and each component has its own smooth mean field across the sky, with an intrinsic covariance about that field chosen so that its total spread equals the fitted covariance; in the second (P), the two components are built from the stars assigned to each.
Every mock catalog was fitted with $K=3$ and $K=4$ (best of three starts) exactly as the data.

\autoref{tab:null} lists the families of mocks.
None of the 120 single-body mocks with a constant intrinsic covariance and Gaussian or moderately heavy tails (N0, N0c, N0g, N1, and N3) reaches the observed BIC$(K{=}3)-$BIC$(K{=}4)$; their maximum is 34.
Two of the 29 mocks with very heavy tails do, but their multivariate excess kurtosis \citep{mardia1970} is 1,900 and 3,400 (the median of their family is 117; \autoref{tab:null}), against 2.4 for the data.
A single body whose intrinsic covariance varies across it and two overlapping populations drawn from the fitted Gaussians both come close to the observed value: 2 of the 20 strip mocks (N0s), none of the 20 grid mocks (N0s2), and 5 of the 20 two-population mocks (P2) reach it.
In the P2 mocks, the two input populations differ by 29~\kms\ in internal proper motion; measured from hard members in shared 1.5$\degr$ cells, the difference is 44~\kms\ (median): hard assignment exaggerates the differences between overlapping components.
The control built from the stars assigned to each component (P) is over-separated by construction, because its input groups were defined by hard assignment.

\begin{deluxetable}{llrrrrr}
\tablecaption{Null-test mock catalogs \label{tab:null}}
\tablehead{
\colhead{Family} & \colhead{Inner body} & \colhead{$N$} & \colhead{Median} & \colhead{Range} & \colhead{$\geq$ data} & \colhead{Kurtosis}
}
\startdata
N0 & one body; Gaussian; quadratic mean field & 30 & $-54$ & $[-95, -21]$ & 0 & 1.6 \\
N0c & one body; Gaussian; cubic mean field & 20 & $-36$ & $[-66, -11]$ & 0 & 1.7 \\
N0g & one body; Gaussian; mean field in $1.5\degr$ cells & 20 & $-9$ & $[-53, +34]$ & 0 & 1.8 \\
N1 & one body; Student-$t$, $\nu=5$ & 30 & $-16$ & $[-67, +28]$ & 0 & 17.7 \\
N3 & one body; Student-$t$, $\nu=30$ (tails as in the data) & 20 & $-44$ & $[-74, -15]$ & 0 & 2.5 \\
N2 & one body; Student-$t$, $\nu=3$ & 29 & $+58$ & $[+29, +1647]$ & 2 & 117.0 \\
N0s & one body; covariance varies in six east--west strips & 20 & $+115$ & $[+78, +182]$ & 2 & 1.9 \\
N0s2 & one body; covariance varies on a $3\times3$ grid & 20 & $+89$ & $[+18, +139]$ & 0 & 2.4 \\
P2 & two populations from the fitted Gaussians & 20 & $+124$ & $[+68, +231]$ & 5 & 1.5 \\
P & two populations from the hard-labeled stars & 10 & $+629$ & $[+579, +694]$ & 10 & 2.2 \\
S22 & whole catalog drawn from the fitted $K=4$ model & 20 & $+162$ & $[+115, +216]$ & 13 & 2.8 \\
\hline
Data & & & $+155$ & & & 2.4 \\
\enddata
\tablecomments{$N$ is the number of mock catalogs (one N2 mock failed to converge and is excluded). Median and range are of BIC$(K{=}3)-$BIC$(K{=}4)$; ``$\geq$ data'' counts the mocks that reach the observed value. Kurtosis is the median multivariate excess kurtosis \citep{mardia1970} of the inner-body residuals. In S22 the whole catalog, not only the inner body, is drawn from the fitted model.}
\end{deluxetable}

\end{document}